\documentclass[lettersize,journal]{IEEEtran}

\usepackage{amsmath,amsfonts}
\usepackage{amssymb}
\usepackage{physics,amsmath}
\usepackage{array}
\usepackage{bm}
\usepackage{mathrsfs}
\usepackage{siunitx}
\usepackage{ifthen}
\usepackage{mathtools}
\usepackage{relsize}

\usepackage{amsthm}
\usepackage{csquotes}
\usepackage{caption}

\usepackage{subcaption}
\usepackage{graphicx}
\usepackage{mathtools}
\usepackage{xcolor}
\usepackage{tikz}
\usetikzlibrary{chains}
\usepackage{makecell}
\usepackage{textcomp}
\usepackage{stfloats}
\usepackage{lipsum}  
\usepackage{bm}
\usepackage{url}
\usepackage{verbatim}
\usepackage{graphicx}
\usepackage{cite}
\usepackage{pdfpages}
\usepackage{tikz-3dplot}
\usepackage{relsize}
\usetikzlibrary{positioning, arrows.meta}
\usepackage{xcolor}
\usepackage[hidelinks]{hyperref} 

\hypersetup{
    colorlinks=false, 
    linkbordercolor={1 0 0}, 
    citebordercolor={0 1 0}, 
    pdfborder={0 0 1} 
}

\usepackage{tikz}
\usetikzlibrary{3d}
\usetikzlibrary{arrows.meta} 

\newcommand{\vect}[1]{\mathbf{#1}}

\usepackage{courier} 
\usepackage{algorithm} 
\usepackage{algpseudocode} 
\usepackage{comment}

\title{Multi Ratio Shift Keying (MRSK) Proposal Draft}

\date{August 2024}

\begin{document}

\newcommand{\orcidiconBak}{\href{https://orcid.org/0009-0004-3410-2176}{\includegraphics[scale=0.1]{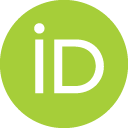}}}

\newcommand{\orcidiconOba}{\href{https://orcid.org/0000-0003-2523-3858}{\includegraphics[scale=0.1]{figures/orcidID128.png}}}

\title{
Multi Ratio Shift Keying (MRSK) Modulation for Molecular Communication
}
\author{Boran A. Kilic,\orcidiconBak,
O. B. Akan,\orcidiconOba~\IEEEmembership{Fellow,~IEEE}
\thanks{The authors are with the Center for neXt-generation Communications (CXC), Department of Electrical and Electronics Engineering, Ko\c{c} University, Istanbul 34450, Turkey (e-mail: boran.kilic@std.bogazici.edu.tr, akan@ku.edu.tr).}
\thanks{Ozgur B. Akan is also with the Internet of Everything (IoE) Group, Electrical Engineering Division, Department of Engineering, University of Cambridge, Cambridge, CB3 0FA, UK (email: oba21@cam.ac.uk).} \thanks{This work was supported by the AXA Research Fund (AXA Chair for Internet of Everything at Ko\c{c} University)}
}

\maketitle

\begin{abstract}
Molecular Communication (MC) leverages the power of diffusion to transmit molecules from a transmitter to a receiver. A wide variety of modulation techniques based on molecule concentration, type, and release time have been extensively studied in the literature. In this paper, we propose a novel modulation technique that encodes the information into the relative concentrations of multiple molecules called Multi Ratio Shift Keying (MRSK) designed for diffusion-based MC without drift. We show that leveraging all possible ratios in a set of molecules can help mitigate the effects of intersymbol interference (ISI) and provide a flexible communication channel. To evaluate the performance of the MRSK, we develop a mathematical framework for studying the statistics of the ratio of random variables, focusing on noncentral Gaussian distributions. We then assess MRSK performance both analytically and through particle-based simulations under various channel conditions, identifying potential sources of error in our system model. Additionally, we conduct a comparative analysis of commonly used modulation schemes in the literature based on bit error rate (BER). The results show that MRSK significantly outperforms all traditional modulation schemes considered in this study in terms of BER. MRSK offers a promising, flexible, and more reliable communication method for the future of the MC paradigm.
\end{abstract}


\section{Introduction}\label{Indtroduction}
\IEEEPARstart{M}{olecular} communication is a highly promising paradigm for realizing nano-networks, due to its biologically inspired approach and close resemblance to communication mechanisms found in nature. Unlike electromagnetic communication systems, MC uses molecules as carriers of information, making it particularly suitable for medical and biological applications, including intrabody nano-networks for continuous health monitoring and targeted drug delivery \cite{MC_medicine_apps}, \cite{fundamentals_of_MC}. Among the various methods explored, diffusion-based MC has been widely studied. However, due to the inherent randomness of diffusion, such systems often face significant challenges, especially ISI, impacting reliability.

Over the past decade, a range of modulation schemes have been introduced, primarily encoding information into three key molecular properties: concentration, molecule type, and release time. Modulation techniques such as On-Off Keying (OOK) \cite{OOK}, Concentration Shift Keying (CSK) \cite{CSK}, Pulse Amplitude Modulation (PAM) \cite{PAM}, Molecule Shift Keying (MoSK) \cite{MOSK}, and Release Time Shift Keying (RTSK) \cite{RTSK} have been widely explored and extensively studied in the literature. In this work, we depart from conventional techniques and focus on the novel concept of encoding information via relative ratios of different carrier molecules, which offers several distinct and practical advantages over traditional schemes.

Research suggests that complex stimuli, such as odors, may be recognized in a concentration-invariant manner, relying primarily on ratio information. For example, studies have shown that rats are capable of distinguishing binary odor mixtures based on the molar ratios of the components, allowing the recognition of mixtures even at varying concentration levels \cite{odor_ratio_rats}. In metabolic networks, the relative concentrations of enzymes and substrates can dictate the direction and rate of biochemical reactions. In glycolysis, the balance between ATP and ADP ratios (rather than their absolute quantities) affects regulatory enzyme activity, controlling the pathway's overall flux. Additionally, when the diffusion coefficients of molecules are similar or nearly identical, their spatiotemporal distribution remains consistent, which is particularly beneficial for maintaining communication integrity under rapidly changing conditions. Furthermore, ratio-based MC allows the same ratio combination to be achieved with any number of molecules, providing flexibility in energy consumption, a critical advantage in resource-limited environments.

To the best of our knowledge, the concept of ratio-based molecular communication is first introduced in Isomer Ratio Shift Keying (IRSK) \cite{6708565}, where it is presented as an extension of concentration-based modulation using isomers as messenger molecules. However, the statistical analysis in this study is limited. A more detailed statistical model is later developed in Ratio Shift Keying (RSK) for ligand binding in both stationary and mobile molecular communication systems \cite{RSK}. Nevertheless, these studies do not consider the impact of channel memory, assuming sufficiently large intervals. Additionally, in both works, encoding is limited to the ratio of two types of molecules, meaning only a single ratio is utilized to carry information. We argue that the effectiveness of ratio-based modulation increases as the number of molecule types used increases and introducing a general M-ary scheme presents great flexibility for different channel conditions. This is especially relevant in high data rate scenarios where symbol time can be extended by encoding more information in each symbol. Moreover, the statistical properties of the ratio of Gaussian random variables have not yet been thoroughly investigated in the context of molecular communication.

We propose a novel modulation scheme, Multi Ratio Shift Keying (MRSK), which encodes information into all possible ratios of a set of different molecules. To assess its performance, we develop a robust mathematical framework that models the received ratio of molecules on the basis of transmitted ratio values. MRSK demonstrates several advantages over other modulation schemes under specific channel conditions. Firstly, it enables the use of a fixed threshold detection mechanism that is independent of channel conditions. The independence from the channel impulse response (CIR) arises from the fact that the expected ratio of the total number of received molecules remains consistent between the transmitter and receiver. Thus, expected value of the ratio is constant in space and time given that molecules have the same diffusion coefficient.

The independence of CIR also makes MRSK especially well-suited for Single Input Multiple Output (SIMO) communication systems. Since MRSK is unaffected by channel conditions, a transmitter can send messages to multiple receivers simultaneously without needing to adjust modulation and demodulation process for each receiver. For instance, molecular communication systems with a central transmitter, such as \cite{bilgen2024mycorrhizalfungiplantsymbiosis}, can employ non-coherent detection across various receivers using MRSK, ensuring efficient molecular utilization. Additionally, MRSK exhibits greater resilience to ISI and Gaussian noise under certain configurations. This resilience arises from the fact that probability density function (PDF) of the ratio of Gaussians has lighter tail compared to Gaussians formed using the same number of molecules. Additionally, the division operation may provide stability by partially mitigating ISI effects. While MRSK enables high transmission rates by encoding extensive information per symbol, it may require more molecules than conventional modulation techniques. Therefore, it is particularly well-suited for application requiring high-precision, in environments with abundant molecular resources.

The rest of this paper is structured as follows. In Section \ref{sytem_model_ratio_of_rvs}, we introduce the physical Single-input Single-output (SISO) system model. Then, we develop a mathematical framework to study the ratio of noncentral Gaussian random variables. In Section \ref{MRSK}, we explicitly demonstrate how the transmission and detection mechanism work while presenting the system parameters. Afterwards we propose three different detection methods explicitly developed for MRSK. In Section \ref{BER}, we derive an analytical expression of BER for two of the proposed detection methods. In Section \ref{Performance Evaluation}, we evaluate the performance of MRSK under various channel conditions both analytically and by simulations. We also compare MRSK with other modulations in the literature in order to prove its superiority under certain channel conditions. Finally, in Section \ref{Conclusion}, we summarize the results and discuss some future directions. 

\section{System Model and Ratio of Random Variables} \label{sytem_model_ratio_of_rvs}
\subsection{General System Topology}\label{sytem_topology}
The system model considered in this paper, similar to those in \cite{6807659} and \cite{7390096}, involves a single transmitter block and a single receiver block in an unbounded 3D MC channel in which molecule motion is only governed by diffusion. This environment is completely filled with a fluid of viscosity $\eta$ exempt from drift currents. For our purposes, the transmitter is modeled as a point source, which is capable of releasing any given number of molecules instantaneously to the channel. The transmitter consists of three main units: molecule reservoirs, a modulator, and a mixing chamber.  The molecule reservoirs separately store $N$ different molecules until they are needed. Modulator is able to precisely select the number of molecules based on the encoding of the information and transports them to a mixing chamber where molecules are homogeneously mixed and released into the channel in the form of concentration spikes. We neglect the volume transmitter device displaces and assume that it does not interfere with the propagation of IM's as the physical design is not of our concern in this paper. 

In a typical MC scenario, the concentration of emitted molecules is significantly lower than that of the fluid molecules. As a result, interactions between the emitted molecules (such as collisions and electrostatic forces) can be disregarded, and the movement of each molecule can be modeled using Brownian motion \cite{brownian_motion}. Given that each molecule moves independently, molecular diffusion follows Fick’s laws of diffusion, with a uniform diffusion coefficient across space The concentration of molecules are articulated as 
\begin{equation}
    \frac{\partial C(x,y,z,t)}{\partial t} = D \ \nabla^2 C(x,y,z,t),
\end{equation}
\begin{equation}
    C_n(x, y, z, t) = \frac{Q_{n}}{\left(4\pi t D\right)^{3/2} } \cdot e^{-{\frac{\abs{\vect{r}}^2}{4Dt}}} ,
    \label{eq:concentration}
\end{equation}
where  $D$ is the diffusion coefficient of molecules for the given medium, $C_n$ defines the spatiotemporal concentration function, $Q_{n}$ denotes the number of molecules that are released from transmitter in $n$-th symbol interval and $\abs{\vect{r}}$ is the Euclidean distance from the receiver.

We assume an absorbing receiver model for the detection of information molecules (IM). Absorbing receiver model degrades every molecule that hits its surface. Receiver is perfectly synchronized with the transmitter, and has full information of transmission times. Absorbing receiver implements \cite{7008529} energy-based detection, which counts the total number of absorbed molecules in each symbol interval. Receiver is a spherical entity whose center is located at a distance $d$ away from the transmitter and its radius is $r$. 

\begin{figure}
    \centering
    \includegraphics[width=\linewidth]{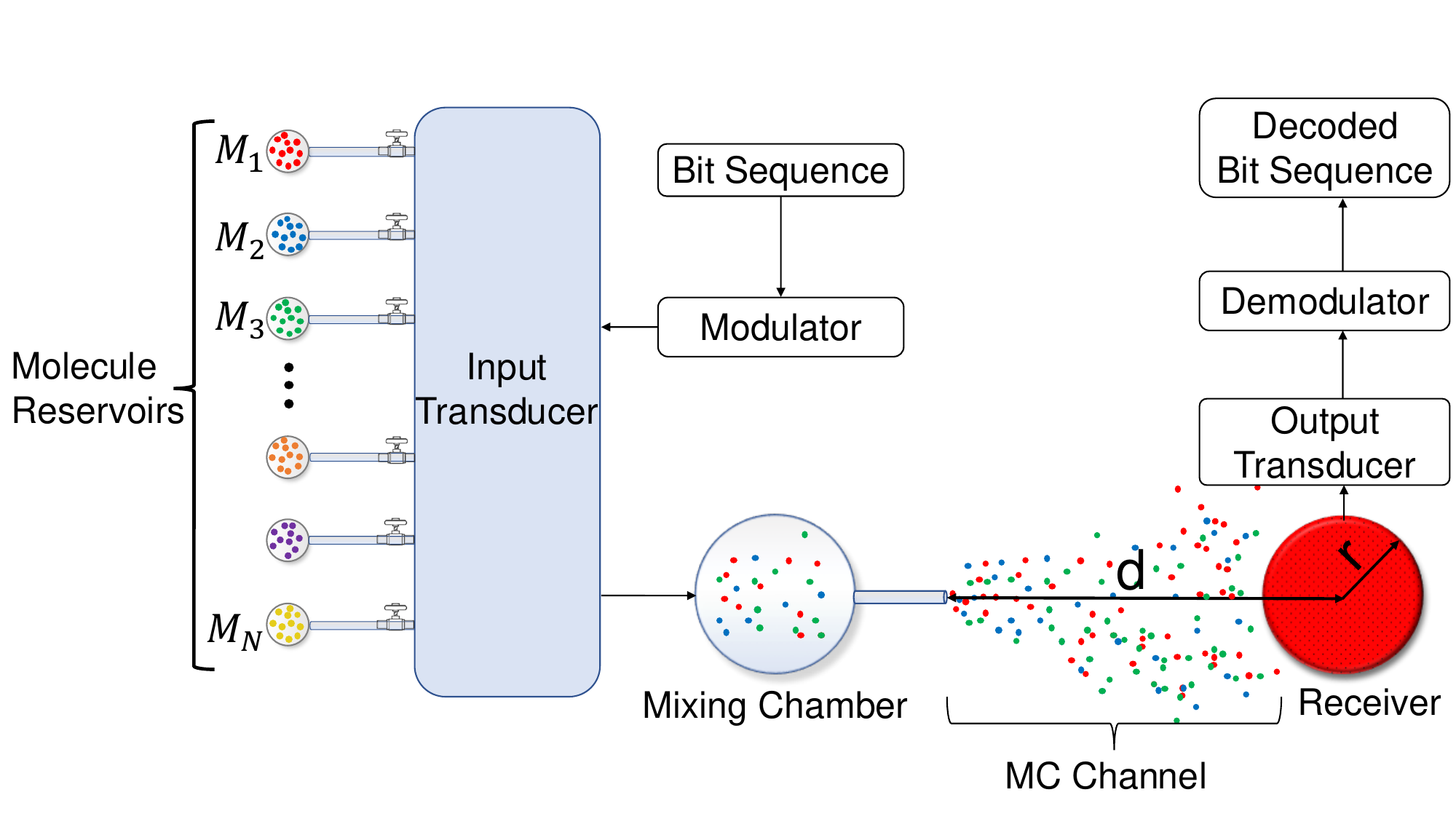
    }\caption{General systems topology of MRSK. Different colors represent different molecules.}
    \label{fig:system_schematic}
\end{figure}

Therefore, the problem at hand comes down to the Fick's unbounded 3D-diffusion equation with initial and boundary conditions as given in \cite{6807659}. The fractional hitting rate of molecules to the receiver can be calculated as 
\begin{equation}
    n_{\text{hit}}(t)= \frac{r}{d}\frac{d-r}{\sqrt{4 \pi D}t^{3/2}}e^{-\frac{(d-r)^2}{4Dt}}.
\end{equation}
By integrating $n_{hit}(t)$ from the release of molecules, $t=0$, to an arbitrary time $t$, we find the fraction of molecules absorbed until time $t$ as
\begin{equation}\label{hit_probability}
N_{\text{hit}}(t) = \frac{r}{d} \, \text{erfc}\left[\frac{d - r} {\sqrt{4Dt}}\right] \ ,
\end{equation} 
where $\text{erfc}(z) = \frac{2}{\sqrt{\pi}} \int_z^\infty e^{-y^2} \, dy$ is the complementary error function.
Notice that the fraction of molecules hitting the receiver by time $t$ is equal to the probability of any given molecule, out of all molecules, hitting the receiver by time $t$.
\begin{equation}
    N_{\text{hit}}(t) = F_{\text{hit}}(t) \ ,
\end{equation}
where $F_{\text{hit}}(t)$ denotes the cumulative distribution function (CDF) of the number of molecules absorbed. Then, a molecule is absorbed by the receiver at the $k$-th interval with probability
\begin{equation}\label{eq:channel_response}
    P_{\text{hit}}(k) = F_{\text{hit}}(k T_s) - F_{\text{hit}}([k-1]T_s ) \ ,
\end{equation}
where $T_s$ is the signaling interval length or inter-symbol time.

Propagation of a single molecule in the channel can be analyzed as a Wiener Process \cite{wiener_process}. Then, total number of molecules absorbed in the $k$-th signaling interval when only 1 molecule is transmitted becomes a Bernoulli random variable. When multiple independent molecules are emitted, the number of molecules absorbed at the k-th interval is a binomial random variable with a success probability of $P_{hit}(k)$. Both Poisson and Gaussian approaches are widely used in the literature to model arrival processes in diffusion-based MC. However, as shown explicitly in \cite{DBLP:journals/corr/YilmazCTP14} Gaussian modeling results in lower root mean square error (RMSE) for a higher number of molecules which aligns with the conditions of our modulation scheme. Assuming the success probability is low enough, total number of molecules detected in the k-th signaling interval is
\begin{equation}\label{binomial_to_gaussian_arrival}
    N_{Rx}[k] \sim \mathcal{N}(\mu[k], \sigma^2[k])  \ ,
\end{equation}
where $\mu[i]$ and $\sigma[i]$ are the mean and variance of this process. 

Due to the causal nature of MC, a portion of the molecules absorbed in a signaling interval may originate from molecules released in previous intervals. This characteristic of MC channel gives rise to ISI, a major issue across all modulations. As the findings of \cite{6807659} suggest, the memory of the channel has a heavy-tailed and infinite nature. However, for all practical purposes, the channel can be modeled with a finite impulse response (FIR) model, by considering only the first $L$ memory elements as presented in \cite{energy_req}. Then, the mean and variance of received number of molecules are
\begin{equation}\label{eq:expected_num_of_mols}
    \mu[k] =  \sum_{i=1}^{L} P_{hit}[i]\ s[k - i + 1] ,
\end{equation}
\begin{equation}\label{eq:standart_dev_of_mols}
   \sigma^2[k] =  \sum_{i=1}^{L} P_{hit}[i]\ (1 - P_{hit}[i])\ s[k - i + 1] ,
\end{equation}
where L is channel memory length or ISI window and $s[i]$ is the transmitted signal, i.e., number of molecules emitted from the transmitter in the beginning of the i-th signaling interval. Note that we exclude the contribution of stationary noise arising from the detection process or channel imperfections, as it significantly complicates the calculations in subsequent steps while adding little value to our analysis.

\subsection{Ratio of Gaussian Random Variables}\label{sec:ratio_of_RV}
MRSK scheme encodes the information in the ratio of number of molecules. In order to elaborate on the distribution of ratio values, we need to construct a mathematical framework that explains the probability distribution of the ratio of two uncorrelated Gaussian random variables. 

Assume we have two independent Gaussian random variables $X$ and $Y$ with means  $\mu_x$, $\mu_y$ and, standard deviations $\sigma_x$, $\sigma_y$, respectively. Also let the ratio of these random variables be $\eta = \frac{X}{Y}$. The region $\mathrm{d}\Omega(\eta)$ is the set of all the points $(x,y) \in \mathbb{R}^2$ bounded between $x = \eta y$ on the one side and $x = (\eta + d\eta)y$ on the other. Every point of this set corresponds to ratio from interval $(\eta, \eta + d\eta)$ or, in other words \cite{solid_approx}:

\begin{equation}
\forall (x,y) \in [\mathrm{d}\Omega(\eta) \subseteq \mathbb{R}^2] : \eta < \frac{x}{y} < \eta + d\eta.
\end{equation}

Note that the Gaussian approximation allows negative $X$ and $Y$ values to have nonzero probabilities. In that sense Gaussian approximation is not perfect since it is physically not possible to receive a negative number of molecules. Nevertheless, this limitation is a minor issue that typically resolves with a sufficiently large mean number of molecules, provided that the channel response is not too small. That is 
\[
s[i] \gg 1 \hspace{0.2cm} \And \hspace{0.2cm} P_{hit}(k) \nsim 0 \implies F_X(0) \ll 1 \ ,
\]
where $F_x$ is the CDF of the given Gaussian distribution.
Therefore, precise formulation of the region $\mathrm{d}\Omega(\eta)$ requires the following set of inequalities:

\begin{equation}
    \begin{aligned}
        \mathrm{d}\Omega(\eta) = \Big\{ (x,y) \in & \mathbb{R}^2 \Big|  \left[ \eta y < x < (\eta + d\eta) y \land y > 0 \right] \\
        & \lor \left[ \eta y > x > (\eta + d\eta) y \land y < 0 \right] \Big\}.
    \end{aligned}
\end{equation}

\begin{figure*}[ht]
    \centering
    \begin{subfigure}[t!]{0.32\textwidth} 
              \includegraphics[width=\columnwidth,keepaspectratio]{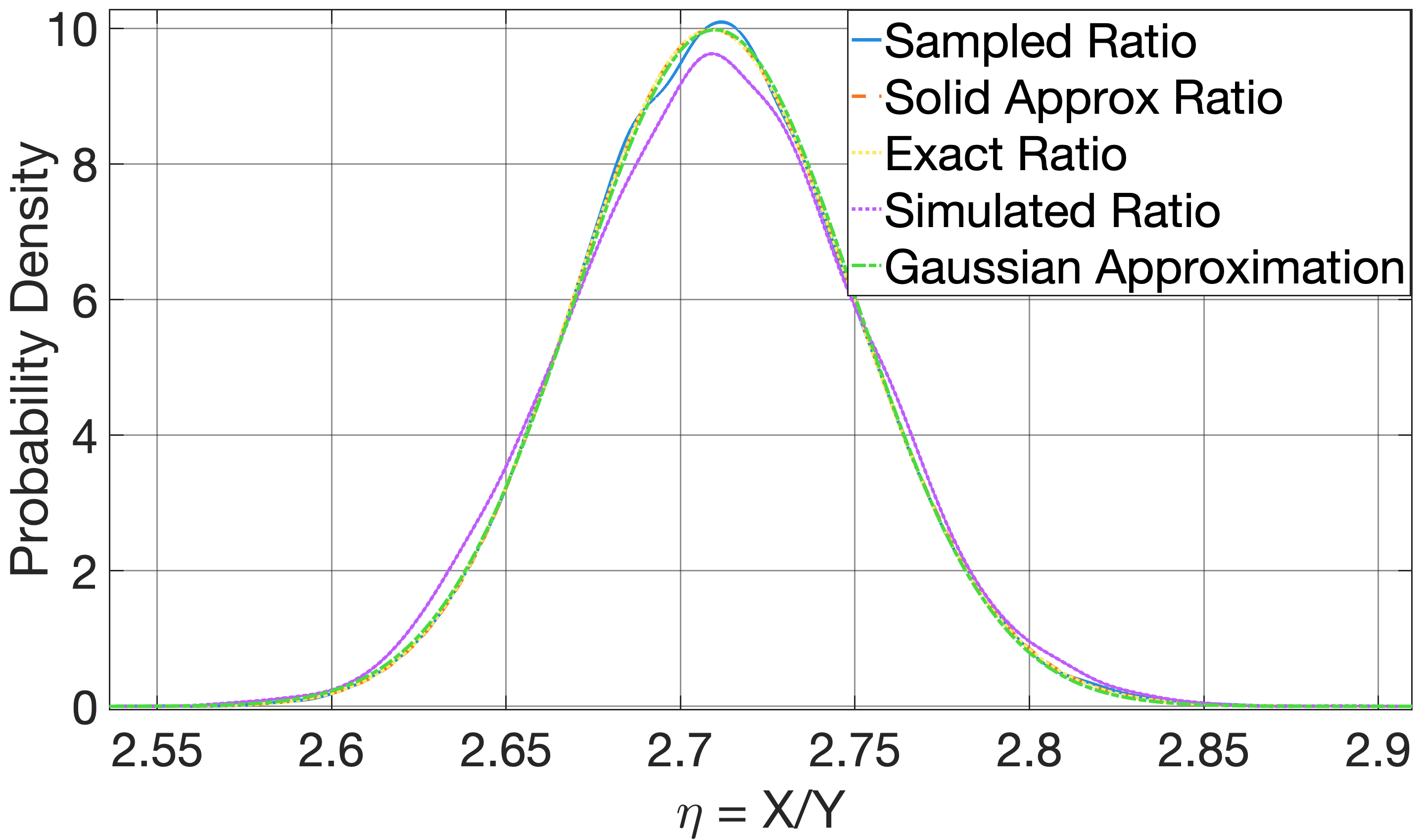} 
        \caption{ratio = $e$}
        \label{fig:subfig_a}
    \end{subfigure}
    \hspace{0.1em}
    \begin{subfigure}[t!]{0.32\textwidth} 

        \includegraphics[width=\columnwidth,keepaspectratio]{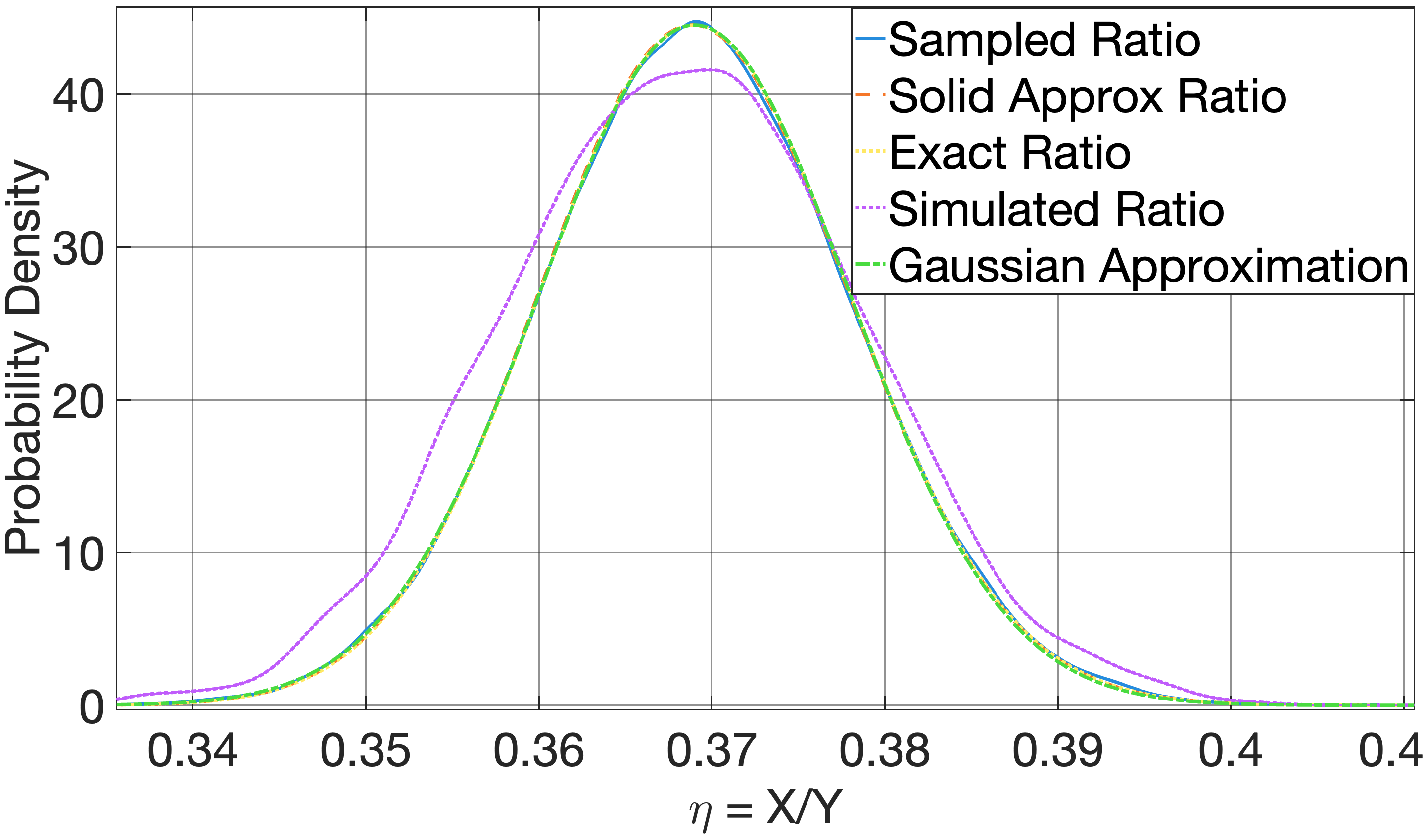} 
        \caption{ratio = $e^{-1}$}
        \label{fig:subfig_b}
    \end{subfigure}
    \hspace{0.1em}
    \begin{subfigure}[t!]{0.32\textwidth} 
        \includegraphics[width=1\columnwidth,keepaspectratio]{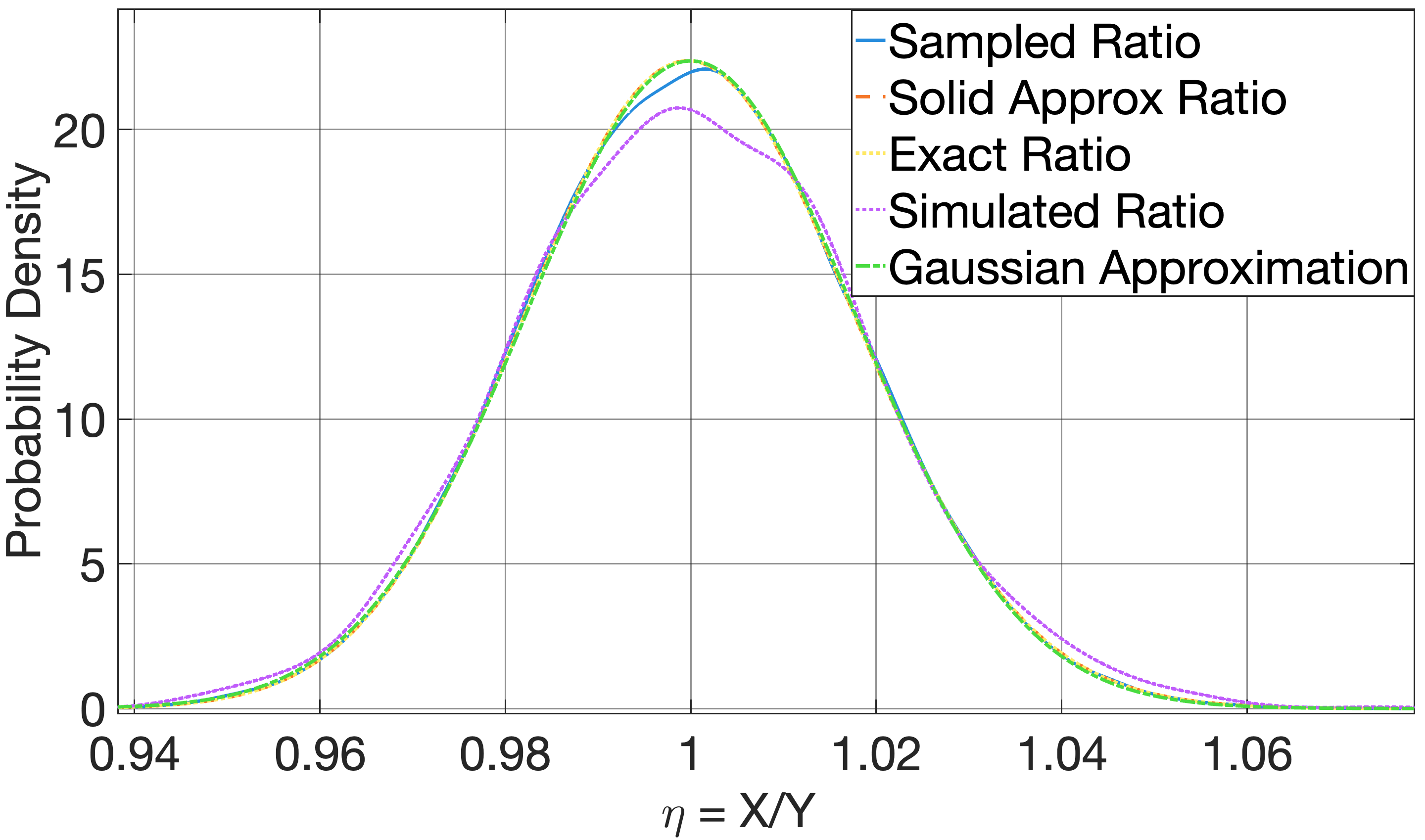} 
        \caption{ratio = 1}
        \label{fig:subfig_c}
    \end{subfigure}
    \caption{Plots of PDFs of the received ratio when transmitted ratios are $e$, $e^{-1}$ and $1$ . }
    \label{fig:ratio_pdf_plots}
\end{figure*}

Furthermore, let $\psi : \mathbb{R} \to [0, \infty)$ be a continuous, non-negative, Riemann-integrable function and $d\eta \to 0^+$ denote an infinitesimal differential of the ratio. Then we say that $\psi$ is the ratio PDF if the following conditions are satisfied:
\begin{itemize}
    \item $P\left(\eta \leq \frac{X}{Y} < \eta + d\eta\right) = \int_\eta^{\eta + d\eta} \psi(t) dt = \psi(\eta) d\eta,$
    \item $\int_{-\infty}^{\infty} \psi(\eta) d\eta = 1.$
\end{itemize}
Accordingly, connection between ratio PDF and Gaussian random variables can be given as
\begin{equation}
    \int\int_{\mathrm{d}\Omega(\eta)} \rho_X(x) \rho_Y(y) dx dy = \psi(\eta) d\eta \ ,
\end{equation}
\begin{equation}
    \psi(\eta) = \int_{-\infty}^{\infty} \abs{y} \rho_X(y\eta) \rho_Y(y)dy .
\end{equation}

The derivation of ratio PDF is not further pursued here since remaining steps involve basic calculus. For further insight, we refer readers to \cite{solid_approx} and simply express the ratio PDF as derived in \cite{ratio_pdf}.
\begin{align}\label{eq:exact_ratio}
    \psi(\eta) &= \frac{b(\eta) \cdot d(\eta)}{a^3(\eta)\sqrt{2 \pi \sigma_x \sigma_y}} \nonumber \times \left[ \Phi \left( \frac{b(\eta)}{a(\eta)} \right) - \Phi \left( -\frac{b(\eta)}{a(\eta)} \right) \right] \nonumber \\
    &\quad + \frac{1}{a^2(\eta) \cdot \pi \sigma_x \sigma_y} \cdot e^{-\frac{c}{2}} 
\end{align}
where
\begin{equation}
    \begin{aligned}
        a(\eta) &= \sqrt{\frac{1}{\sigma_x^2}\eta^2 + \frac{1}{\sigma_y^2}}, \ \ \ \ \ b(\eta) = \frac{\mu_x}{\sigma_x^2} \eta + \frac{\mu_y}{\sigma_y^2}, \\
        c &= \frac{\mu_x^2}{\sigma_x^2} + \frac{\mu_y^2}{\sigma_y^2}, \ \ \ \ \ \ \ \ \ \ \   \ d(\eta) = e^{\frac{b^2(\eta) - c \cdot a^2(\eta)}{2a^2(\eta)}} 
    \end{aligned}
\end{equation}
and $\Phi$ is the normal CDF,
$\Phi(t) = \int_{-\infty}^{t} \frac{1}{\sqrt{2 \pi}} e^{-\frac{1}{2}u^2} du$. Although this expression accurately describes the PDF of the ratio of two uncorrelated Gaussian random variables, it is too complex for analytical calculations and it lacks a closed form CDF which is required for BER analysis in Section \ref{BER}. Therefore, we adopt the solid approximation of $\phi(\eta)$ implemented in \cite{solid_approx} for the knot-efficiency problem.

To contruct the solid approximation, we introduce a triplet of new parameters \( p, q, r \) based on a quartet of known parameters \( \bar{x}, \bar{y}, \sigma_x, \sigma_y \in \mathbb{R}^+ \) by the following equations:
\begin{equation}\label{p_q_r}
    p = \frac{\bar{x}}{\sqrt{2}\sigma_x}, \quad q = \frac{\bar{y}}{\sqrt{2}\sigma_y}, \quad r = \frac{\bar{x}}{\bar{y}} \ .
\end{equation}

In \cite{solid_approx}, limits of $p$, $q$, $r$ are strictly defined based on real-world data about knot efficiency and the validity of the solid approximation is shown for this sub-sector. However, we do not have clear limitations on CIR of the diffusive MC systems. As will be apparent in Section \ref{MRSK}, MRSK modulation assumes a predefined set of ratios $x\in{{\Omega^{-1},\ldots,\Omega^1}}$. Validity of the solid approximation should also be confirmed for the given range of $x$ values. Instead of trying to find some limitations on $p$, $q$, $r$ and creating an approximation from scratch, we directly adopt \cite{solid_approx} and check its validity for this specified range of ratios through numerical computation and particle simulations. Solid approximation of the ratio PDF is given by 
\begin{equation}\label{eq:solid_approx}
    \psi^{\dagger}(\eta) = \frac{p}{\sqrt{\pi} \, \operatorname{erf}(q)} 
    \frac{1 + \frac{p^2 \eta}{q^2 r}}{\left(1 + \frac{p^2 \eta^2}{q^2r^2} \right)^{3/2}} 
    e^{-\frac{p^2 (\frac{\eta}{r}-1)^2}{1 + \frac{p^2}{q^2}\frac{\eta^2}{r^2}} },
\end{equation}
where $erf(x)=\frac{2}{\sqrt{\pi}} \int_0^x e^{\frac{-t^2}{2}}dt$. The closed form CDF of the solid approximation can then be found by:
\begin{equation}\label{eq:solid_approx_CDF}
\begin{aligned}
    P(\eta < \eta_0) &= \Psi^{\dagger}(\eta_0) = \int_{-\infty}^{\eta_0} \psi^+(\eta) d\eta \\
    &= \frac{1}{2} \left( 1 + \frac{\text{erf} \left( \frac{\frac{p}{r} (\eta_0 - 1)}{\sqrt{1 + \frac{p^2}{q^2} \frac{\eta_0^2}{r^2}}} \right)}{\text{erf}[q]} \right) .
\end{aligned}
\end{equation}

It is possible to further approximate the ratio's PDF as a Gaussian. Under some restrictions on means and standard deviations of $X$ and $Y$, Gaussian approximation of the ratio $\eta=X/Y$ is sufficiently accurate for practical purposes. Distribution of the random variable $\eta$ can be approximated by a Gaussian with $\eta \sim \mathcal{N}(\beta, \lambda^2)$ and PDF is expressed as 

\begin{equation}\label{eq:gaussian_approx}
   \psi^{\dagger\dagger}(\eta) = \frac{1}{\sqrt{2\pi\lambda^2}}\exp[{-\frac{(x-\beta)^2}{2\lambda^2}}],
\end{equation}
where

\begin{equation}\label{eq:gaussian_approx_mu_sigma}
    \beta = \frac{\mu_x}{\mu_y}, \quad 
    \lambda^2 = \frac{\mu_x^2}{\mu_y^2} \left( \frac{\sigma_x^2}{\mu_x^2} + \frac{\sigma_y^2}{\mu_y^2} \right) .
\end{equation}

To verify the validity of our approximations, PDF's of ratio values are plotted in Figure \ref{fig:ratio_pdf_plots}. In this figure, the sampled ratio represents the PDF of two ratio variables directly sampled from a Gaussian distribution and serves as the ground truth for the two approximations. The exact ratio represents the PDF constructed from \eqref{eq:exact_ratio} while the solid approximation is calculated according to \eqref{eq:solid_approx} and the Gaussian approximation is found by \eqref{eq:gaussian_approx}. As can be seen; the PDFs for the exact ratio, solid approximation, and Gaussian approximation are very close to each other for all the ratios $e^{-1},1,e$. Sampled ratio PDF was obtained by taking 10 000 samples from 2 different Gaussians with parameters defined in \eqref{eq:expected_num_of_mols} and observing the distribution of their ratio. Since number of samples is limited, it is not a perfect representation but still in all the cases sampled ratio is very close to the exact and approximated PDF's. The simulation ratios are obtained by repeatedly (8000 times) running a simulation of our system model in a stochastic spatial particle simulator called Smoldyn \cite{smoldyn}.

\section{Multi Ratio Shift Keying} \label{MRSK}
Assume we have a set of $N$ different information molecules (IM). Let the absolute quantities of these molecules be denoted as $Q_1, Q_2, \ldots, Q_N$. Then, there exist $N-1$ ratios, i.e.,
\begin{equation}
    \eta_i = \frac{Q_{i+1}}{Q_i}, \quad 
    \vec{\eta} = \left[ \frac{Q_2}{Q_1}, \frac{Q_3}{Q_2}, \ldots, \frac{Q_{N}}{Q_{N-1}} \right] \ ,
\end{equation}
which we can encode the information in. To construct different symbols, threshold values that define the intervals within the detection range must be established. Let us divide the interval $[-1,1]$ into $2^M-1$ equal sub-intervals where $M$ is the system hyperparameter that determines how many bits each ratio represents. The set of ratio indicators are simply given by
\begin{multline*}
    x' \in X'=\left\{ -1, -1 + \frac{2}{2^M-1}, -1 + 2 \frac{2}{2^M-1}, \ldots, \right. \\
    \left. -1 + \left( 2^M - 2 \right) \frac{2}{2^M-1}, 1 \right\}.
\end{multline*}
The set of predefined ratio values can be given as
\begin{equation}\label{eq:release_ratios}
\begin{aligned}
        x = \Omega^{x'} \in X &= \left\{ \Omega^{-1}, \Omega^{-1 +\frac{2}{2^M-1}}, \ldots, \Omega^{-1 +  \frac{2\left( 2^M - 2 \right)}{2^M-1}}, \Omega^1 \right\}, \\
        x^{(i)} &= \Omega^{-1+\frac{2(i-1)}{2^{M-1}}}, \quad i = 1,2,\ldots,2^{M} .
\end{aligned}
\end{equation}
Clearly, $N-1$ ratios does not fully define the absolute concentrations of $N$ molecules. Therefore, we keep one of the molecules release concentration $Q_1 = Q $ constant as a reference. Then, absolute molecule numbers can be given by the cumulative product of the ratios as
\begin{equation}\label{eq:conc_given_ratios}
    Q_i = Q\prod_{k=1}^{i-1}\eta_k \ .
\end{equation}

Our motivation behind choosing the release ratio values as described is twofold. Firstly, information molecules are scarce in most MC applications, necessitating a restriction on the range of possible ratios. MRSK is not very efficient in terms of molecule consumption, since in the extreme case, the transmitter might need to use $\Omega^{N-1}$ times more of one molecule than another. Therefore, we constrain the range of possible ratios to $[\Omega^{-1}, \Omega^1]$. Henceforth, $\Omega = e$ is used as the default value for the range of ratios due to its natural occurrence and ease of likelihood calculation. We show in Fig. \ref{fig:ratio_pdf_plots} that for this range, PDF approximations closely match the simulated ratio distribution. Nonetheless, the choice of the range of $\Omega$ is arbitrary. Other range options are further analyzed in Section \ref{Performance Evaluation} in terms of BER.

\begin{figure}[t]\label{pdf_plots_of_all_ratios}
    \centering
    \includegraphics[width=0.9\columnwidth]{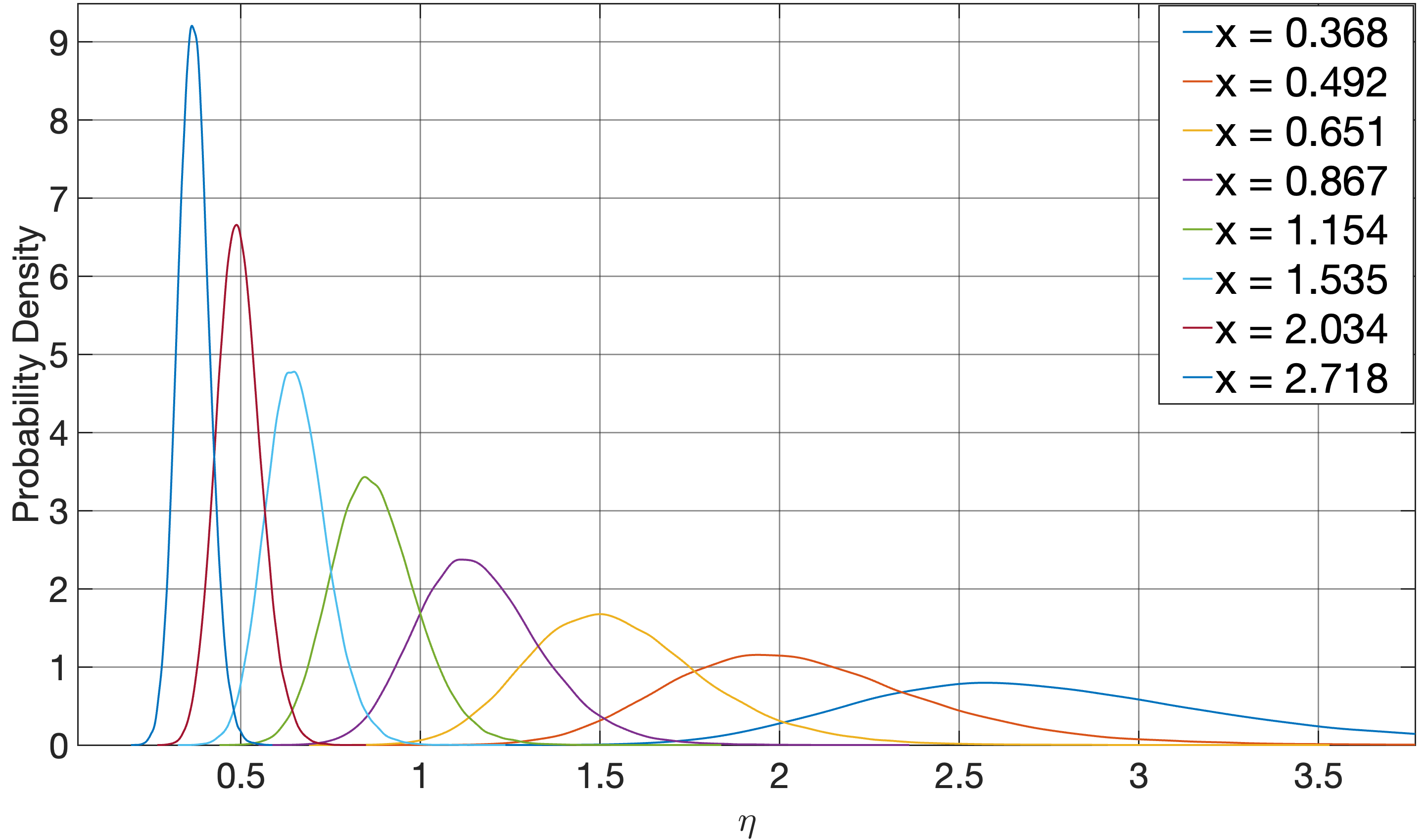}
    \caption{PDF plots of the received ratios for release ratios  $x_{M=3}$ .}
    \label{fig:pdf_plots_M=3}
\end{figure}

Secondly, as evident from \eqref{eq:gaussian_approx}, the standard deviation scales linearly with the release ratio, i.e., the expected value of the received ratio. To address this scaling effect, the intervals are exponentiated, which spreads the threshold values more for higher $\eta$ values, enhancing the receiver's robustness to errors. To validate this approach, we plot the PDFs of all ratios in the set $x_{M=3}$ in Fig. \ref{fig:pdf_plots_M=3}. These PDFs are obtained using the solid approximation for the ratio of two Gaussian random variables \eqref{eq:solid_approx}, with parameters $Q=1000$, $T_s=1s$, $r=5$, and $d=10$. As the mean ratio value increases, the corresponding standard deviation also increases. By placing the predetermined ratio values exponentially, the PDFs are spread more evenly, ensuring that the $\eta$ values with equal probabilities of belonging to two adjacent PDFs are positioned midway between the peaks, as illustrated in Fig. \ref{fig:pdf_plots_M=3}. A similar method is employed in M-ary Concentration Shift Keying, where an arbitrary base $\Gamma$ is used instead of $e$ \cite{M-ary}.
\subsection{Transmission}
We have N types of molecules at our disposal and we set the number of different ratio values we can utilize to be $2^M$. At each transmission interval, encoded communication molecules are released into the channel simultaneously. Then we can represent $(2^M)^{N-1}$ different symbols in each transmission interval which corresponds to $M(N-1)$ bits. This configuration achieves a higher transmission rate than any modulation proposed in \cite{tranmitter_and_receiver_arch}.

To encode the information to be transmitted,we divide the bit sequence into parts of length $M(N-1)$ bits. Each of the M bits in this sequence encodes one particular ratio. 
\begin{equation}\label{eq:encoded_symbol_decimal}
    O_i = [b_{(i-1)M},b_{iM}]_{10} + 1\to  x'_i = -1+\frac{2(O_i-1)}{2^M-1},
\end{equation}
where $O_i$ is the decimal value of the i-th M bits or the symbol that represents ratio $i$.

\begin{figure}[th]
    \centering
    \includegraphics[width=0.8\columnwidth]{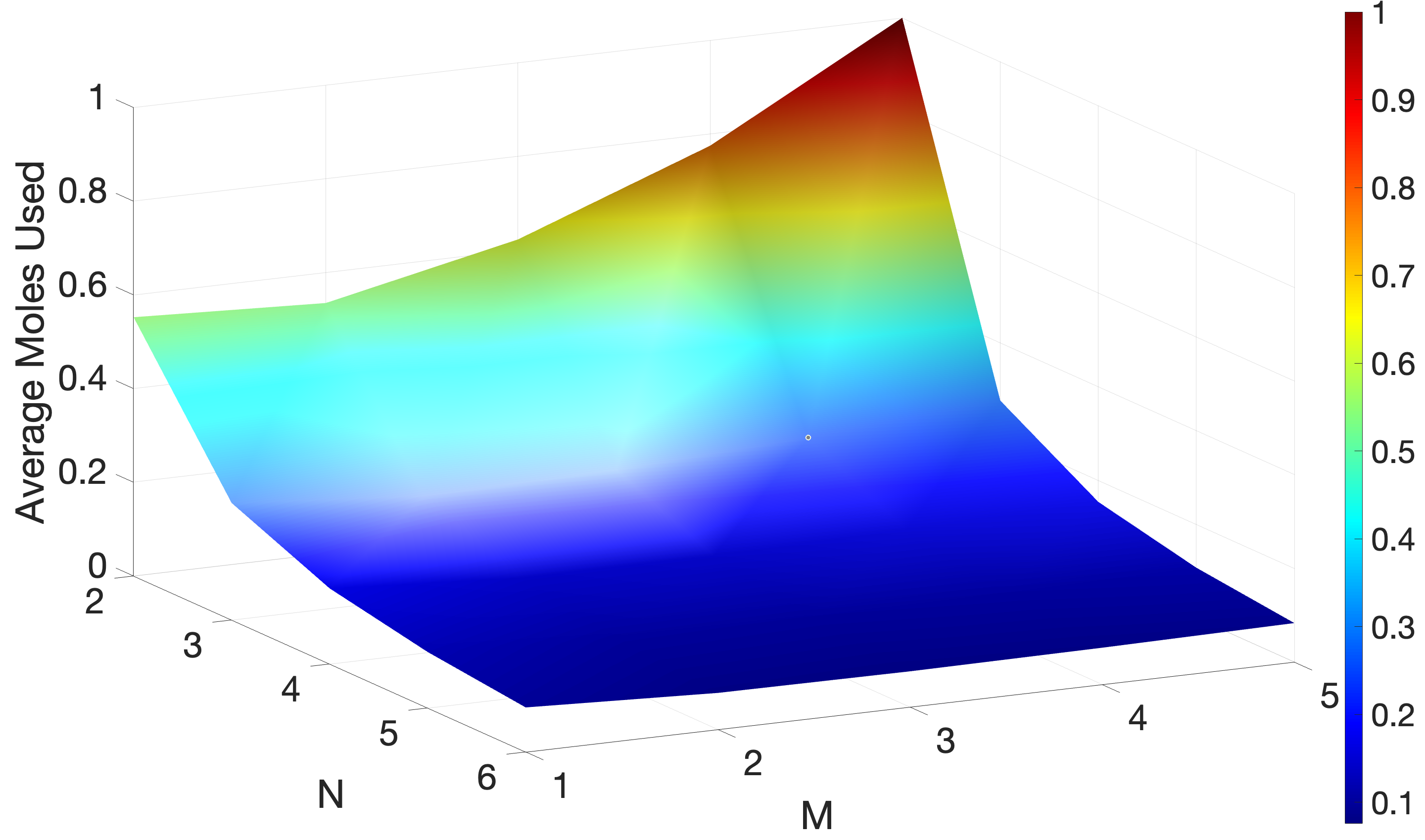}
    \caption{Normalized average number of molecules used per bit in MRSK for a range of $N$, $M$ parameters.}
    \label{fig:ave_mols_used}
\end{figure}

Nano-machines usually operate in environments where the sources are scarce. Although proposed system model is equipped with reservoirs of $N$ different types of molecules, it will eventually run out of molecules since it has a high average molecule consumption as shown in Figure \ref{fig:ave_mols_used}. Especially for large values of $M$, the mean number of transmitted molecules scales exponentially while $N$ counteracts this effect. Depending on the message nano-machines transmit, some bits may occur more frequently than others. This means certain molecules are used more than others. In order to maintain longest possible operation time, it is more than preferable to manage the molecules to be transmitted effectively. This can be easily achieved by interchanging the molecular composition of ratios. 
In detection, some errors may arise from either uncertainty inherent in Gaussian arrivals or previously transmitted molecules. In general, incorrectly detected ratios are more likely to correspond to adjacent symbols, as the probability of detecting a ratio decreases the further it deviates from the expected value. This is a direct result of the fact that PDF of ratios are unimodal as we observe in Section \ref{sec:ratio_of_RV}. Even though Gray coding has no impact on the symbol error probability (SEP), Gray coding can minimize the BER by ensuring that misclassified symbols likely have only a single bit error. Although for $M=1$ encoding stays the same, it ensures optimal source coding for $M>1$. Luckily there is no alteration that needs to be done to the system model except a simple encoding and decoding block.
\subsection{Detection}
Absorbing receiver in our system utilizes energy-based detection, which records the total number of molecules absorbed by the receiver during a prespecified symbol interval $T_s$. After receiver counts the total received number for each different molecule, demodulator calculates each ratio according to the pilot signal formed by the adaptive transmission algorithm outlined in the previous section. Received signal on the k-th signaling interval depends on the current and past $L-1$ transmissions.  We adopt the solid approximation \eqref{eq:solid_approx} due to its rather simpler PDF expression compared to the exact ratio distribution \eqref{eq:exact_ratio} and the existence of a closed form CDF to express the probability distribution of the received ratio.

Substituting $p=\frac{\mu_i}{\sqrt{2}\sigma_i}$  , 
$q=\frac{\mu_{i-1}}{\sqrt{2}\sigma_{i-1}}$, $r=\frac{\mu_i}{\mu_{i-1}}$ into the solid approximation PDF \eqref{eq:solid_approx}, probability of detecting the i-th ratio at the k-th symbol interval $z_i[k]$ given the past $L$ transmitted ratios $\underline{\mathbf{x}}[k-L+1:1]$ is given by
\begin{equation}
    \begin{split}
        f(z_i[k] \ \mid \ \underline{\mathbf{x}}[k&-L+1:k]) = \frac{\mu_i[k]}{\sqrt{2\pi} \, \ \mathrm{erf}\left(\frac{\mu_{i-1}[k]}{\sigma_{i-1}[k]}\right)} \\
        &\times \frac{\mu_{i-1}[k] \ \sigma_i^2[k] + \mu_i[k] \ \sigma_{i-1}^2[k] \ z_i[k]}{(\sigma_i^2[k] + \sigma_{i-1}^2[k])^{3/2}} \\
        &\times \exp \left( -\frac{(\mu_{i-1}[k] \ z_i[k] - \mu_i[k])^2}{2 (\sigma_i^2[k] + \sigma_{i-1}^2[k] \ z_i^2[k])} \right) .
    \end{split}
\end{equation}
Combining \eqref{eq:conc_given_ratios} with \eqref{eq:expected_num_of_mols} and  \eqref{eq:standart_dev_of_mols} we obtain the signal-dependent mean and standard deviation of the i-th molecule in the k-th signaling interval as
\begin{equation}
\begin{aligned}
    \mu_i[k] &= Q \sum_{m=1}^L \prod_{j=1}^i P_{\text{hit}}[m] x_j[k-m+1] ,  \\
    \sigma_i^2[k] &= Q \sum_{m=1}^L \prod_{j=1}^i P_{\text{hit}}[m] (1 - P_{\text{hit}}[m]) x_j[k-m+1] .
\end{aligned}
\end{equation}

Probability of each received ratio remains independent of other received ratios. Therefore, we can write the joint PDF, probability of jointly receiving a set of ratios $z_1[k] = \eta_1, \ z_2[k] = \eta_2, \ldots,  \ z_N[k] = \eta_{N-1} $, of the whole symbol as the product of individual ratios 
\begin{equation}\label{eq:symbol_prob_cond}
    f( \underline{\mathbf{z}}[k] \mid \underline{\mathbf{x}}[k-L+1:k]) =  \prod_{i=1}^{N-1} f( z_i[k]  \mid  \underline{\mathbf{x}}[k-L+1:k]) ,
\end{equation}
where $\underline{\mathbf{z}}[k]=[z_1[k]  \ldots  z_{N-1}[k] ]^T$ is the received ratio vector.

\subsubsection{Fixed Threshold Detection (FTD)}\label{FTD}
In traditional modulation schemes such as CSK, the expected number of received molecules is influenced by various channel properties, including channel length, receiver geometry, and signaling intervals. However, the ratio of received molecules remains independent of these channel characteristics because each molecule is subject to the same channel constraints. It is essential to highlight that this assertion holds due to the linear relationship between the expected number of received molecules and the number of transmitted molecules as evident from \eqref{eq:gaussian_approx_mu_sigma}, particularly when ISI effects are ignored. In this scenario, the channel response described in \eqref{eq:channel_response} does not depend on the number of molecules. This independence arises from the MC paradigm's approach which usually neglects interactions among individual molecules and analyzes the collective behavior of an ensemble of molecules using statistical methods.

The advantage of ratio modulation is that we can determine fixed thresholds based on release ratios, maximizing the probability of correct classification without having access to CIR. CIR independence is crucial in the context of Internet of Bio-Nano-Things \cite{6708565}. In most applications of diffusive molecular communication, nano-machines, along molecules, also experience Brownian motion. Mobility forces CIR to be time-dependent and probabilistic, nullifying the optimality of detection schemes reliant on channel responses. To overcome this problem, we present a detection scheme that uses release ratios $x$ given by \eqref{eq:release_ratios}. An MC model employing ratio-based communication in the case of mobile transmitter and receiver was thoroughly analyzed in \cite{RSK}.

The $i$-th symbol detected in the $k$-th slot, $z_i[k]$,
is obtained by applying the maximum likelihood (ML) decision rule, i.e., 
\begin{equation}
    z_i[k] =  \underset{{x}_i[k]}{argmax} \ f(R_k | H_x) ,
\end{equation}
where $x \in X$ and $f(R_k|H_x)$ denotes the conditional PDF of $R_k$ given that hypothesis $H_x$ holds when ISI is neglected. Assuming $\mu_1<\mu_2<\ldots<\mu_{2^M}$ and  $\sigma_1<\sigma_2<\ldots<\sigma_{2^M}$, which are satisfied in our encoding scheme, ML decision rule reduces to pairwise comparisons around threshold values where $\mu_i$ and $\sigma_i$ are the expected value and standard deviation of the received ratios when predefined ratio $x^{i}$ is transmitted (as defined in \eqref{eq:expected_num_of_mols}, \eqref{eq:standart_dev_of_mols} for $L=0$). If we use the Gaussian approximation \eqref{eq:gaussian_approx} for the ratios of total received number of molecules, decision thresholds can be obtained from

\begin{equation}
\frac{e^{-\frac{(\mathcal{E}_i - \beta_i)^2}{2\lambda_i^2}}}{\sqrt{2\pi \lambda_i^2}}  = 
\frac{e^{-\frac{(\mathcal{E}_i - \beta_{i-1})^2}{2\lambda_{i-1}^2}}}{\sqrt{2\pi \lambda_{i-1}^2}} , \quad i = 1, \dots, M-1
\end{equation}
where
\begin{equation}
    \beta_i = \frac{Q^{(Tx)}_{i+1}}{Q^{(Tx)}_i} \ ,  \quad \lambda_i = \frac{Q^{(Tx)}_{i+1}}{Q^{(Tx)}_i} \sqrt{\frac{1-P_{\text{hit}}[1]}{Q^{(Tx)}_i P_{\text{hit}}[1]} + \frac{1-P_{\text{hit}}[1]}{Q^{(Tx)}_{i+1} P_{\text{hit}}[1]}} \ .
\end{equation}

We adopt the approach in \cite{M-ary_threshold} and approximate the expected value and the standard deviation to find 
\begin{equation}
    \mathcal{E}_i = \sqrt{\beta_{i+1}\beta_i} = \sqrt{x_{i+1}{x_i}} \ .
\end{equation}
Then for $\Omega=e$, the exact thresholds become
\begin{equation}\label{exact_thresholds}
    \mathcal{E}_i = 
        \displaystyle \exp{{-1 + \frac{2i - 1}{2^M - 1}}} \quad,\quad i = 1,2,\ldots, 2^M-1 \ .
\end{equation}
and the decision rule can be expressed as
\begin{equation}\label{eq:FTD}
    \tilde{z}_k = 
    \begin{cases}
        x^{(1)}[k], \quad \text{if} \quad z[k] < \mathcal{E}_1, \\
        x^{(i)}[k], \quad \text{if} \quad \mathcal{E}_{i-1} \leq  z[k] \leq \mathcal{E}_{i} \ ,\\
        x^{(2^M)}[k],   \ \text{if} \quad \mathcal{E}_{2^M-1} \geq z[k]  
    \end{cases}
\end{equation}
where $z_k$ is the received ratio, $\Tilde{z}_k$ is the detected symbol and $x_k^{(i)}$ is the transmitted symbol in the $k$-th signaling interval.

\subsubsection{Adaptive Detection with Memory Cancellation (ADMC)}
FTD calculates ML based on the assumption that the total number of molecules received in the symbol interval $k$ is solely due to molecules transmitted in the current interval. This approach is valid for large $T_s$ when the CIR of previous transmissions is negligible compared to current transmission. However, for applications requiring high data rate, we have to consider ISI. Unfortunately, we cannot utilize ratios to address ISI since a particular ratio does not define the state of a molecule by itself. State of the system depends collectively on all the ratios as apparent in \eqref{eq:conc_given_ratios}.

To effectively mitigate the impact of ISI, it is essential to first estimate the number of molecules originating from previous time slots. By obtaining the PDF of the number of ISI molecules, conditioned on the currently received molecules, the ISI effect can be estimated through the conditional mean of this distribution. Accordingly, we present the approach of memory-$\Gamma$ cancellation that is also adopted by \cite{memory_cancel}. Memory-$\Gamma$ cancellation takes into account ISI effect by subtracting the estimated contribution of the past $\Gamma$ transmissions assuming past $\Gamma$ symbols were received correctly.

For practical channel conditions, majority of the ISI can be attributed to the last transmitted symbol and contributions from earlier signaling intervals can approximately be neglected.  Additionally, nano-machines in MC paradigm have limited buffer capacity making it difficult to retain the number of all recently received molecules for each $2^M$ molecules, even temporarily in memory. For the reasons stated and due to its simplicity, we adopt 1-memory cancellation. 
More specifically, the adjusted total received number of molecules is
\begin{equation}\label{eq:ADMC}
    \begin{split}
         Q_i^{\text{Rx-adapted}}[k] &= Q_i^{\text{Rx}}[k] - P_{\text{hit}}[2] \, \tilde{Q}_i^{\text{Tx}}[k-1], \\
        \tilde{Q}_i^{\text{Tx}}[k-1] &= Q \prod_{j=1}^{i-1} z_j[k-1]
    \end{split}
\end{equation}
where $Q_i^{\text{Rx}}[k]$ is the number of received molecules, $Q_i^{\text{Rx-adapted}}[k]$ is the adapted molecule number and $\tilde{Q}_i^{\text{Tx}}[k-1]$ is the estimated number of transmitted molecules in the previous signaling interval based on the detection decision of the receiver. Although this detection technique is dependent on CIR, MC-Rx does not need exact channel conditions. It only requires $P_{hit}[2]$ to operate. In that sense, adaptive detection can still be utilized unless variations in relative positions of MC-Tx and MC-Rx is severe. 

After adapted total number of received molecules are found, symbols are detected using FTD given in the previous Section \ref{FTD}. A vulnerability of adaptive detection is the fact that success relies on the assumption that previously transmitted symbols are classified correctly. If a previous symbol is misclassified, it can affect later symbols and error can propagate in a cumulative fashion. However, this is rarely the case and ADMC has a tendency to equalize received ratios.

\subsubsection{ML sequence detection}

The previous detection schemes are not by any means optimal because they did not consider the contribution of whole channel memory to the current symbol PDF. In order to choose the most probable symbol we need to set the decision rule so that it maximizes the posterior probability given all of the previous transmissions. ML sequence detection determines the sequence of transmitted ratios ${{x_i}}$ by maximizing the likelihood, i.e.,  
\begin{equation}
    \underline{\mathbf{\hat{x}}}[k-L+1:k] = \underset{\underline{\mathbf{x}}[k-L+1:k]}{argmax}  f(\underline{\mathbf{z}}[k-L+1:k] \mid  \underline{\mathbf{x}}[k-L+1:k]) .
\end{equation}
Here, $\underline{\mathbf{z}}[k-L+1:k]$  and $\underline{\mathbf{z}}[k-L+1:k]$ are $N \times L$ matrices denoting the sequence of transmitted and received ratios, respectively, during each interval from symbol $k-L+1$ to $k$ where $L$ is the ISI length assuming state of the system at any time depends on the most recent $L$ inputs. A priori probabilities of transmitting a certain predefined ratio $x^{(i)}$ are equal across all ratios $x_i$ and probability of transmitting a symbol is equal to
\begin{equation}
    f(\underline{\mathbf{x}}[k]) = \frac{1}{2^{M(N-1)}} \ .
\end{equation}
Therefore, ML sequence detection is equivalent to MAP detection for MRSK. Similar to the branch metric of the trellis search proposed for OOK \cite{ML_detection}, branch metric of  MRSK for the ML criterion can be given by taking the logarithm of the probability of the solid approximation PDF, i.e.,

\begin{equation}
    \begin{split}
        \mathcal{M}_i^{\dagger-\text{ML}}[k] &= \displaystyle -\ln \left[ \mathrm{erf} \left( \frac{\mu_{i-1}[k]}{\sigma_{i-1}[k]} \right) \right]  \\
         & \ \ \  -\frac{3}{2} \ln \left( \sigma_i^2[k] + \sigma_{i-1}^2[k] z_i^2[k] \right) \\
        & \quad - \frac{\left( \mu_{i-1}[k] z_i[k] - \mu_i[k] \right)^2}{2\left( \sigma_i^2[k] + \sigma_{i-1}^2[k] z_i^2[k] \right)} + \ln \left( \mu_i[k] \right) \\
         &\quad + \ln \left( \mu_{i-1}[k] \sigma_i^2[k] + \mu_i[k] \sigma_{i-1}^2[k] \right) 
    \end{split}
\end{equation}
This is the branch metric of the $i$-th ratio in the $k$-th symbol interval. Current mean and standard deviation of the Gaussian arrivals $\mu_i[k]$, $\sigma_i[k]$ can be derived using past $L$ transmissions $\underline{\mathbf{x}}[k-L+1:k]$. One can get a simpler but less accurate metric by using the Gaussian approximation  given in \eqref{eq:gaussian_approx}.

\begin{equation}
    \begin{split}
    \mathcal{M}_i^{\dagger\dagger-ML}[k] &= -2 \ \ln{\left(\frac{\mu_i[k]}{\mu_{i-1}[k]} \ \sqrt{\frac{\sigma_i^2[k]}{\mu_i^2[k]} + \frac{\sigma_{i-1}^2[k]}{\mu_{i-1}^2[k]}} \ \right)} \\
    &\quad -  \frac{ \mu_{i-1}^2[k] \ \left( \ \mu_{i-1}[k] z_i[k] - \mu_i[k] \right)^2 \ }{2\left( \sigma_i^2[k] \ \mu_{i-1}^2[k]+ \sigma_{i-1}^2[k] \ \mu_i^2[k] \ z_i^2[k] \right)} 
    \end{split}
\end{equation}
We observe that each ratio in the current symbol is independent of each other, meaning probability of receiving a certain ratio $z_i[k] = \eta_i$ does not depend on other ratios in the symbol. Moreover, each symbols sent over the transmission history are also independent which makes the conditional PDF
\begin{align}
    &f \big( \underline{\mathbf{z}}[k-L+1:k] \mid  \mathbf{x}[k-L+1:k] \big) = \\
    & \prod_{m=k-L+1}^k f \big( \underline{\mathbf{z}}[m] \mid \mathbf{x}[k-L+1:m] \big) 
\end{align}
where $f(\ \underline{\mathbf{z}}[k] \mid \mathbf{x}[k-L+1:k])$ is as given in \eqref{eq:symbol_prob_cond}. Finally, we can express the ML sequence detection decision rule as

\begin{equation}
    \underline{\mathbf{\hat{x}}}[k-L+1:k] = \underset{\underline{\mathbf{x}}[k-L+1:k]}{argmax} \sum_{j=1}^{N-1}\sum_{m=k-L+1}^{k}  \mathcal{M}_j^{\dagger-\text{ML}}[m] \ .
\end{equation}

Clearly, ML detection is computationally intensive. The ML sequence detector must construct a trellis and calculate the probabilities for all $2^{LM(N-1)}$ possible symbol sequences, resulting in a computational complexity that scales as $\mathcal{O}(2^{LM(N-1)})$. Even for a relatively short channel memory, such as $L=5$, the number of required calculations reaches astronomical levels even for modest values of $N$ and $M$. For instance, setting $N=5$, $M=3$ results in $2^{60} \approx 1.15 \times 10^{18}$ calculations, each involving multiple nonlinear operations. Performing such computations is highly expensive, even for conventional computers, let alone nano-machines. Thus, using ML detection for symbol decoding in MRSK is not feasible unless \(N\) and \(M\) are kept minimal or we opt to sacrifice from accuracy by choosing a relatively low L.

\section{Bit Error Rate analysis} \label{BER}

For assessing the performance of the proposed scheme in which a single symbol error can result in multiple bit errors, BER is chosen as the evaluation metric because, unlike symbol error probability, which focuses on the probability of receiving the wrong symbol, BER measures the fraction of received bits that are erroneous relative to the total number of transmitted bits. BER takes into account the error correction coding schemes that may be applied to the data. Even in cases where the SNR is low and symbol errors are frequent, coding techniques can recover many bit errors, thus improving the BER. Furthermore, BER is versatile and can be applied across all modulations, including binary and higher-order schemes.

The theoretical BER expression requires the calculation of all possible symbol sequences of length $L$ $\mathbf{x}[k-L+1:k]$ and can be given by  
\begin{equation}
    BER = \sum_{\forall \mathbf{x}[k-L+1:k]} \left(\frac{1}{2^{LM(N-1)}}\right) P_e(\mathbf{x}[k-L+1:k]) \ ,
\end{equation}
where $P_e(\mathbf{x}[k-L+1:k])$ is the error probability at the k-th symbol interval given that the sequence $\mathbf{x}[k-L+1:k]$ is transmitted. For FTD error probability can be expressed as 
\begin{equation}\label{eq:BER_FTD}
    \begin{split}
    P_e(\mathbf{x}[k-L+1:k]) = \sum_{j=1}^{N-1} \sum_{i=1}^{2^M} &\frac{\mathbf{d}_{hamming} (O_i , O_{x_j[k]})}{M(N-1)} \\ &\times P(\tilde{z}_j[k] = x^{(i)}) \ .
    \end{split}
\end{equation}
Here, $\mathbf{d}_{hamming}(\cdot)$ is the Hamming distance operator, which takes in two decimal numbers and calculates the number of different bits in their binary representation. $O_i$ represents the decimal symbol number of ratio $i$ given in \eqref{eq:encoded_symbol_decimal} and $O_{x_j[k]}$ is the ground truth for the symbol number of the received ratio $j$ where $O_i,  O_{\tilde{z}_j[k]} \in {1,2,\ldots, 2^{M}}$. 
In each symbol, we calculate the error probability of each ratio $i$ and sum the results. Due to random Gaussian arrivals, each received ratio $j$ has a probability of corresponding to other $2^M-1$ symbols. We calculate the probability that a received ratio is decoded as $x^{(i)}$ and multiply with the number of erroneous bits caused by this misdetection. For each symbol sequence, bit error probability sum is divided by $N(M-1)$ for normalization.

In FTD, in addition to $2^{M-1}$ thresholds defined in \eqref{eq:FTD}, we define $\mathcal{E}_0= -\infty , \mathcal{E}_{2^M}= \infty$ for completeness. One might object that we should set the lower bound to zero since ratio value is non-negative. However, we model the stochastic arrival of molecules as Gaussians, leading to a nonzero probability of receiving a negative number of molecules. Therefore, for formalism, we must consider negative ratios when normalizing the PDF. Then probability of the received ratio $z_j[k]$ being detected as $x^{(i)}$ can be calculated by

\begin{equation}\label{eq:FTD_prob}
    \begin{split}
        P(\tilde{z}_j[k] = x^{(i)}) &= P(\mathcal{E}_{i-1} \leq z_j[k] \leq \mathcal{E}_{i}) \, \\
        &= \int_{\mathcal{E}_{i-1}}^{\mathcal{E}_{i}} \psi(\eta) \, d\eta \ ,
    \end{split}
\end{equation}
where $\psi(\eta)$ is the exact ratio distribution given in \eqref{eq:exact_ratio}. This expression is extremely complex and cannot be calculated analytically. Luckily, we derive the closed form CDF of the solid approximation ratio of 2 uncorrelated noncentral Gaussian random variables \eqref{eq:solid_approx_CDF} which gives very similar results to the exact expression for practical purposes as confirmed in Fig. \ref{fig:ratio_pdf_plots}. Substituting $p=\frac{\mu_j[k]}{\sqrt{2}\sigma_j[k]}$, $q=\frac{\mu_{j-1}[k]}{\sqrt{2}\sigma_{j-1}[k]}$, $r=\frac{\mu_j[k]}{\mu_{j-1}[k]}$ and simplifying, we get 

\begin{equation}\label{eq:symbol_prof_ito_erf}
    \begin{split}
        P(\tilde{z}_j[k] = x^{(i)}) =  \frac{1}{2} &\left[ \frac{\text{erf} \left( \frac{\mu_{j-1} (\mathcal{E}_{i} - 1)}{\sqrt{2 (\sigma_j^2 + \sigma_{j-1}^2 \mathcal{E}_{i}^2)}} \right)}{\text{erf} \left( \frac{\mu_{j-1}}{\sqrt{2} \sigma_{j-1}} \right)} \right. \\
        &- \left. \frac{\text{erf} \left( \frac{\mu_{j-1} (\mathcal{E}_{i-1} - 1)}{\sqrt{2 (\sigma_j^2 + \sigma_{j-1}^2 \mathcal{E}_{i-1}^2)}} \right)}{\text{erf} \left( \frac{\mu_{j-1}}{\sqrt{2} \sigma_{j-1}} \right)} \right] .
    \end{split}
\end{equation}

For ADMC, BER calculation is not so simple. Since thresholds are dynamic, modifying the effective number of received molecules, we have to consider all possible sequences of detected symbols on top of received number of molecules. The calculations are the same up to \eqref{eq:BER_FTD} but $ P(\tilde{z}_j[k] = x^{(i)})$ differs for ADMC. Since the received symbol in the previous interval can correspond to any one of the $2^M$ symbol, probability that $j$-th ratio is decoded as symbol $i$ can be calculated by 
\begin{equation}
    \begin{aligned}
    P(\tilde{z}_j[k] = x^{(i)}) = \sum_{m=1}^{2^M} &\mathcal{P}(\tilde{z}_j[k] = x^{(i)}\ |\ \tilde{z}_j[k-1] = x^{(m)}) \\ &\times\ P(\tilde{z}_j[k-1] = x^{(m)}) \ ,
    \end{aligned}
\end{equation}
where $\mathcal{P}(\tilde{z}_j[k] = x^{(i)}\ |\ \tilde{z}_j[k-1] = x^{(m)})$ is the probability of detecting the $j$-th ratio as $x^{(i)}$ when previous symbol was detected as $x^{(m)}$. $\mathcal{P}(\tilde{z}_j[k] = x^{(i)}\ |\ \tilde{z}_j[k-1] = x^{(m)})$ can be calculated as in \eqref{eq:FTD_prob}, but $p$, $q$, $r$ values in \eqref{eq:symbol_prof_ito_erf} are now calculated using the adapted concentrations where the contribution of the previously transmitted symbol is excluded, given by \eqref{eq:ADMC}. In order to calculate $ P(\tilde{z}_j[k-1] = x^{(m)})$, we go one step back and calculate the probability in the $(k-1)$-th symbol interval in the same manner. We follow this procedure, by iteratively calculating probabilities just like going down the nodes of a M-ary search tree \cite{search_tree} until we arrive at the base node which corresponds to the first symbol interval. Carrying out a similar calculation for every possible branch of this search tree requires enormous computation time and memory, especially as $M$ increases. Strictly speaking, ADMC requires  $2^{M(L+1)}-2^M$ times more calculations than FTD. Runtime required to evaluate ADMC analytically quickly reaches astronomical levels and therefore, we will evaluate the BER performance of ADMC via simulations only.

\section{Performance Evaluation}\label{Performance Evaluation}

In this section, we evaluate the effect of various channel and system parameters on the BER and conduct a comparative analysis of the proposed MRSK modulation scheme against conventional MC baseline schemes such as OOK, CSK, MoSK, and RTSK. The theoretical BER is computed using \eqref{eq:BER_FTD}. To validate these theoretical calculations, we use a stochastic particle-based simulation tool, Smoldyn \cite{smoldyn}, which allows rule-based configuration of MC systems. As described in Section \ref{sytem_topology}, a predefined number of molecules, determined by the transmitted message, is released from a point source. The molecules undergo Brownian motion after emission, and a user-defined spherical receiver counts those within its boundaries. The molecules are then removed from the simulation environment, representing full absorption at the receiver. Neither physical saturation \cite{7895189} of the receiver nor molecule re-emission into the channel was considered in our model or simulations since since it is beyond this paper's scope. Particle simulations do not perfectly mimic MC systems, as they discretize time into small intervals and may overlook events between time steps. Generally, this issue can be mitigated by using sufficiently small time steps. However, due to the large number of possible symbol sequences in MRSK, we had to set relatively larger time steps to complete simulations within a reasonable time frame.

To fairly assess the performance of MRSK and other modulation schemes, we normalize the bit time $t_b$, ensuring that error rates can be compared between different settings at the same data transmission rate. For MRSK, symbol time is $T_s = M(N-1)t_b$. The hannel parameters are fixed at $d=10\mu m$, $r=5\mu m$, $D=79.4\mu m^2/s$, $Q=1000$, $L=5$, $t_b=0.5$s $M=1$ for varying $M$ and $N=2$ for varying $N$, unless otherwise specified.
\begin{figure}[b]
    \centering
    \includegraphics[width=\linewidth]{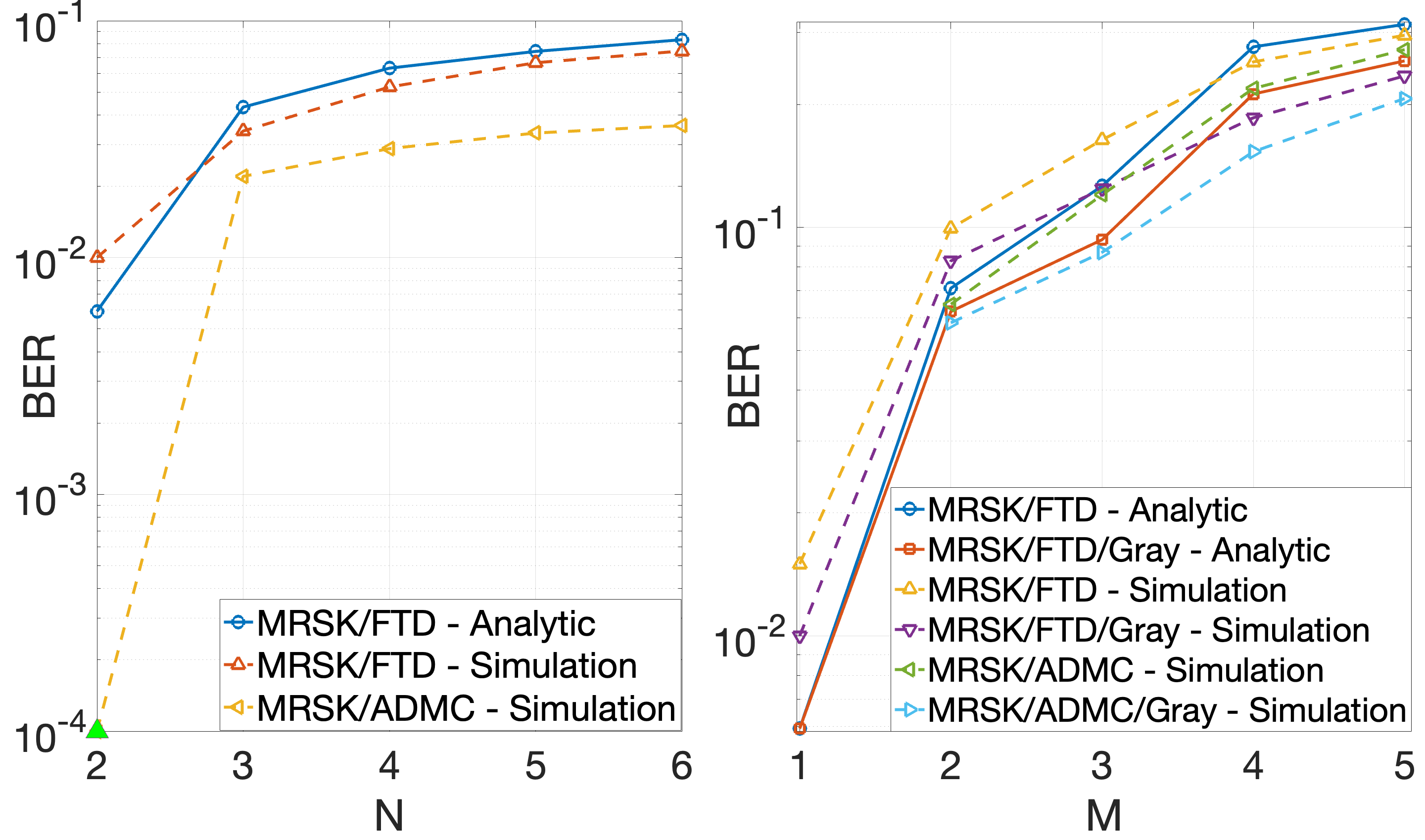}

    \caption{BER vs $N$ and $M$ curves for $t_b=0.5$s .}
    \label{fig:MRSK_varying_N_M_tb0.5}
\end{figure}

\subsection{Effect of M and N Values}
As introduced in Section \ref{MRSK}, number of types of molecules $N$ and modulation order $M$ are single-handedly the two most important since they determine the type of modulation. Accordingly, we first determine the optimal $M$, $N$ values for our system and then analyze the effect of channel conditions on the best performing $M$, $N$ configuration.
\begin{figure}[t]
    \centering
    \includegraphics[width=\linewidth]{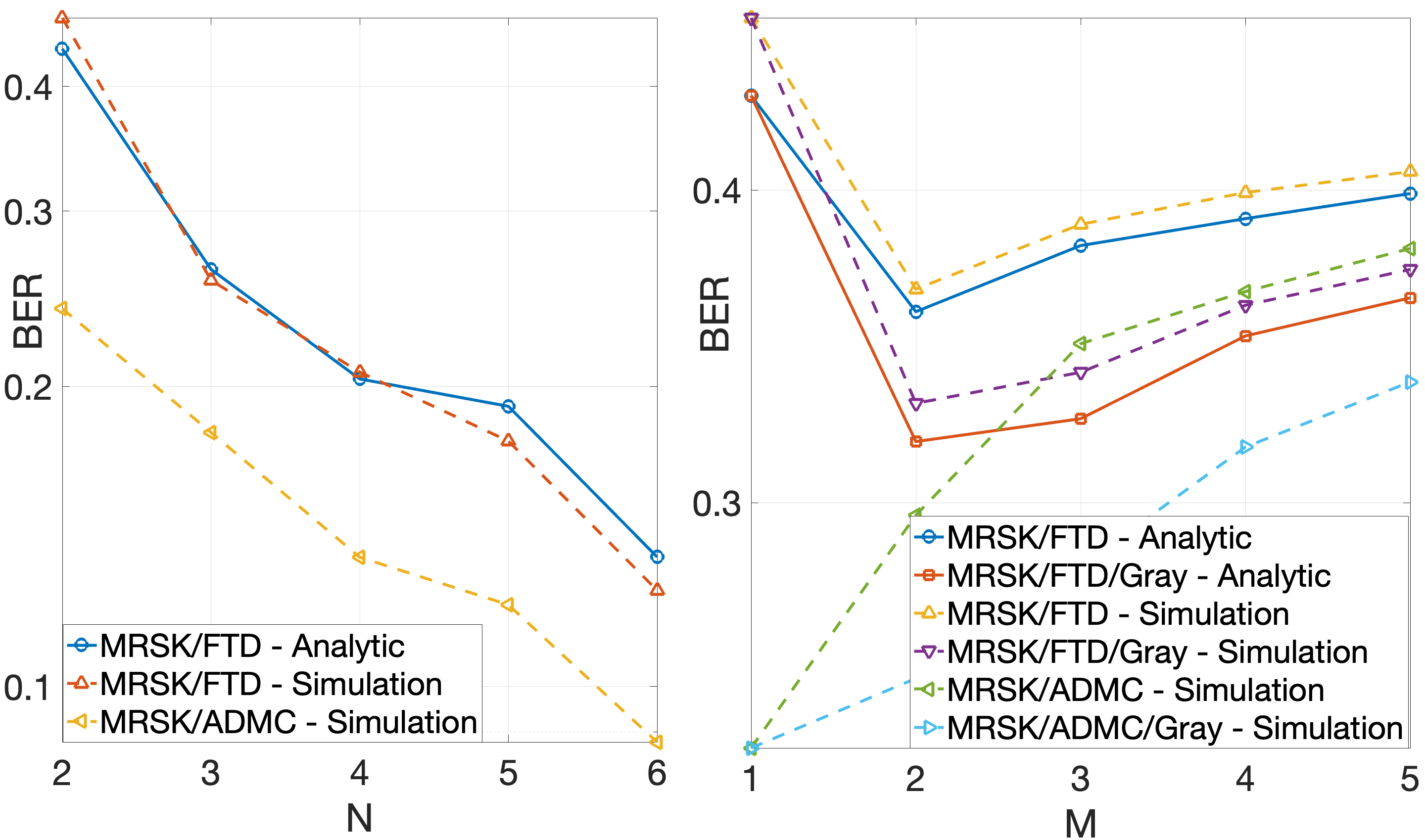}
    \caption{BER vs $N$ and $M$ curves for $t_b=0.05$s .}
    \label{fig:MRSK_varying_N_M_tb0.05}
\end{figure}

In Figure \ref{fig:MRSK_varying_N_M_tb0.5}, the BER is given for $N$ from 2 to 6. BER increases as $N$ increases since the number of transmitted particles of $i$-th molecule $Q_{i}$ depends not only on the ratio $x_i$ but all prior ratios $x_{j<i}$. This means uncertainty in the Gaussian noise from prior ratios accumulates, and make the decision of $i$-th ratio more difficult leading to more errors. Channel length is set to $L=3$ for this subsection due to limited number of simulation runs bounded by hardware capacity available, but it reverts back to $L=5 $ for the remainder of our analysis. Even after thousands of simulation runs, no errors were observed for ADMC at $N=2$, which indicates that the BER is orders of magnitudes lower. The green triangle in $N=2$ is just a placeholder.

Similarly, in Figure \ref{fig:MRSK_varying_N_M_tb0.5}, the BER for varying $M$ values are plotted under the same channel conditions and $N=2$. As $M$ increases or the range is divided into more intervals, the PDFs belonging to adjacent ratios become more intertwined. Consequently, the tails of the distribution \enquote{leak out} to adjacent symbol PDFs and this results in a higher chance of misdetection as observed in Fig. \ref{fig:pdf_plots_M=3}. Accordingly, binary scheme performs significantly better, especially for $N=2$. In addition, Gray coding improves the BER performance for both FTD and ADMC since symbols that are more likely to be misdetected are only one bit away in the binary space.

Two main sources of error in our system are Gaussian noise due to the inherent randomness in diffusion and ISI caused by previously transmitted molecules. One becomes more significant than the other based on channel conditions and type of modulation. Increasing symbol time or transmitter receiver distance increases the gap between subsequent channel responses, effectively dampening the effect of previously transmitted symbols. On the other hand, increasing $N$ results in a higher variance for the ratio of the number of molecules received and increasing $M$ causes adjacent ratio PDFs to have a larger intersection which beclouds the decision process. However, since the bit time is normalized across different channel conditions and $N$,$M$ values; increasing $M$ and $N$ can also decrease ISI by increasing $T_s$. This key observation justifies why we might require to use more than 2 molecules, hence the name \enquote{multiple ratios}.

Although increasing $M$ and $N$ seems to deteriorate performance, we present a contradictory result. For applications requiring high data rate, the bit time may be too short. Increasing $M$ or $N$ can compensate for this limitation by extending symbol time to a moderate level. We plot the BER performance for varying $M$, $N$ values, reducing $t_b$ to $10\%$ of its former value ($t_b =0.05$) in Figure \ref{fig:MRSK_varying_N_M_tb0.05}. BER decreases with increasing $N$ and $M$ until reaching a minimum and starts to rise again. Prior to this minimum, ISI is so dominant that increasing $N$, $M$ decreases BER since symbol time is extended to moderate levels. As $N$, $M$ further increases, ISI effect becomes less significant and BER performance mainly suffers from Gaussian noise. Note that when $N$ is varied in Figure \ref{fig:MRSK_varying_N_M_tb0.05}, minimum is observed at $N=5$ for ADMC while FTD continues to decrease and minimum is likely to be located in the sequel. This is because ADMC already takes precaution against ISI by subtracting the estimated number of molecules transmitted in the previous interval from current number of received molecules. Also minima for varying $M$ values is at $M=2$, and BER gradually increases after this point which shows $M$ is more closely linked to Gaussian noise than $N$. 
\begin{figure}[ht]
    \centering

    \centering
    \includegraphics[width=0.5\textwidth]{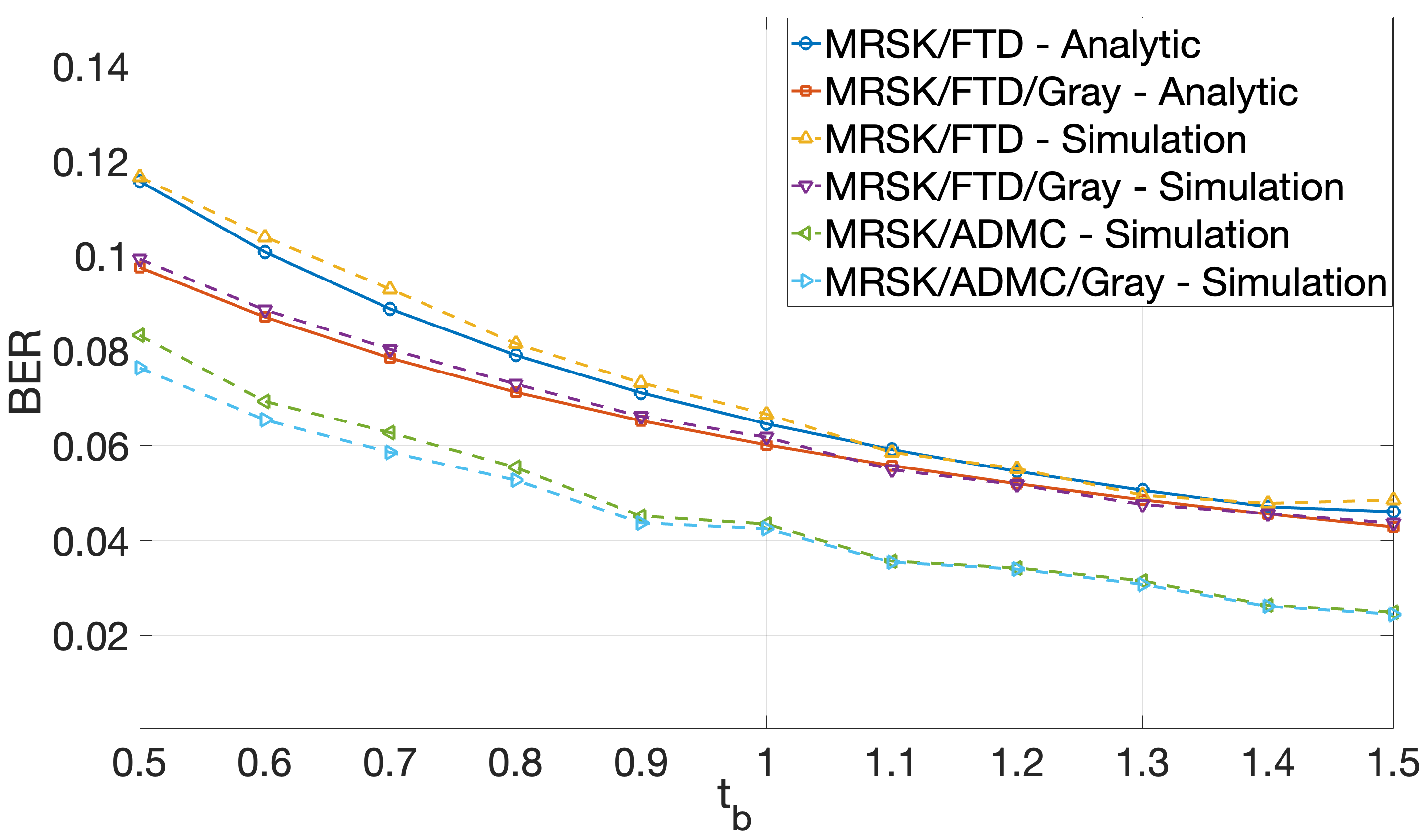} 
    \caption{BER vs $t_b$ curves of MRSK.}
    \label{fig:MRSK_varying_tb}
\end{figure}
\subsection{Effect of Channel Conditions}

\subsubsection{Bit Time}\label{bit_time}

To analyze the effect of bit rate constraint, BER performance for FTD and ADMC with varying $t_b$ are presented in Figure \ref{fig:MRSK_varying_tb}. As expected, BER of MRSK decreases with increasing the bit time and mitigates the effect of ISI \cite{ASSAF201740}. In conformity with previous results, Gray coding and ADMC detection enhance detection performance. We also see that BER of binary and Gray coding converge as $t_b$ increases. 

Although the ISI is less significant with longer inter-symbol time, we have another effect in play that sets a bound for both types of source coding. When the receiver waits for an extended period to sample, molecules from previous transmissions diffuse over a larger region of space that encompasses the receiver. Over time, these molecules accumulate in this region, resulting in a high-concentration environment around the receiver. As a consequence, the receiver appears to uniformly absorb a fixed number of molecules during each signaling interval. This phenomenon effectively saturates the receiver, reducing its sensitivity to molecules from the current transmission and introducing an unavoidable error floor. Both binary and Gray coding approaches eventually converge to this error floor. It is important to emphasize that this saturation is not related to the physical capacity of the absorbing receiver but rather refers to the diminished signal sensitivity.

\begin{figure}[t!]
    \centering
    \includegraphics[width=\linewidth]{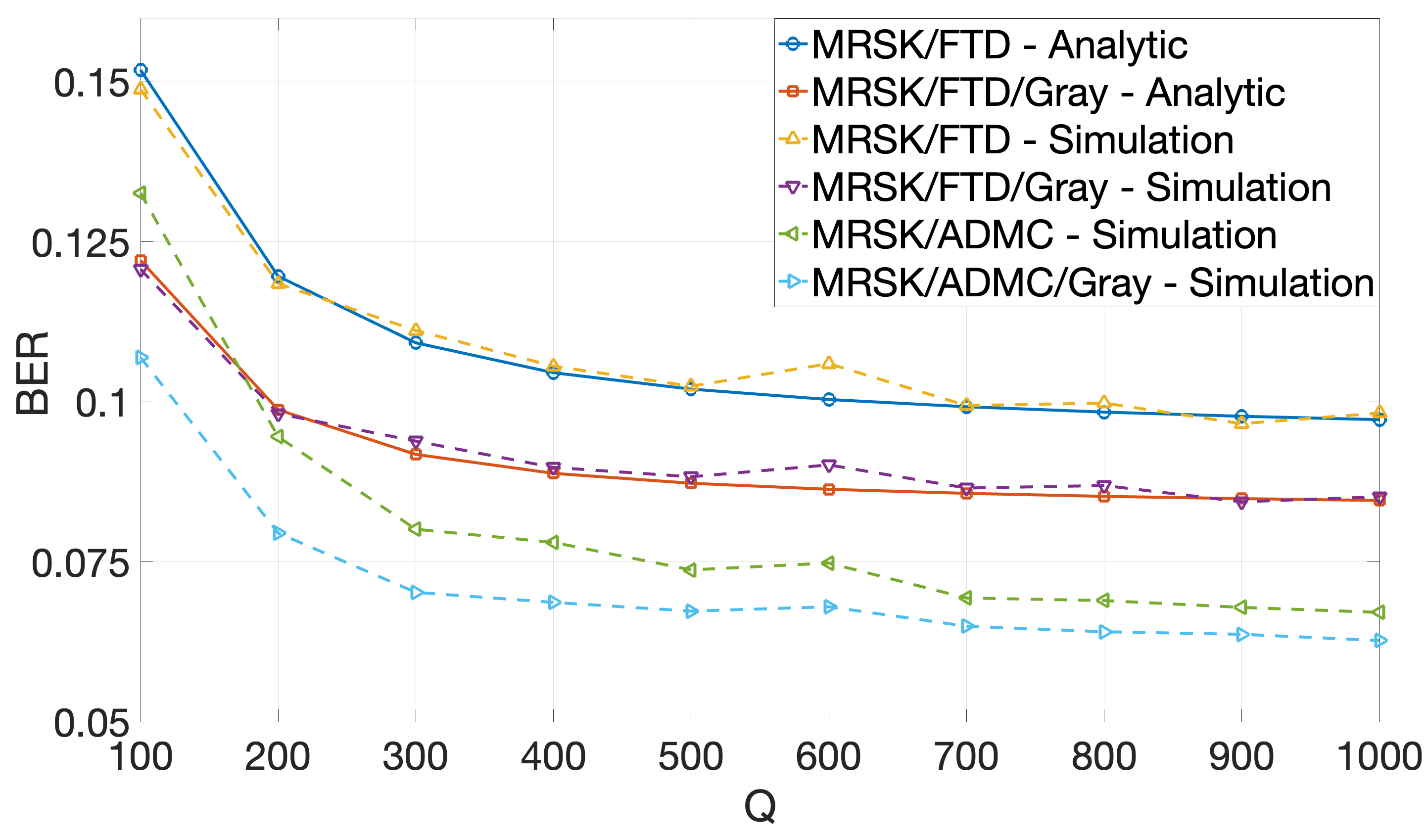} 
    \caption{BER vs $Q$ curves of MRSK.}
    \label{fig:MRSK_varying_Q}
\end{figure}
\subsubsection{Base Number of Molecules}
The number of each transmitted molecule is directly related to our system parameter $Q$ which is the base number of molecules presented in Section \ref{MRSK}. To see the effect of molecule number in BER performance, we vary $Q$ from 100 to 1000 molecules. As seen in Figure \ref{fig:MRSK_varying_Q}, analytic calculations closely match simulation results and the BER decreases as $Q$ increases for both FTD and ADMC. As can be seen from \eqref{eq:gaussian_approx_mu_sigma} the expected value of ratio PDF is proportional to $\frac{\sigma_{x}^2}{\mu_{x}^2}$ and we know, from \eqref{eq:expected_num_of_mols}, that expected value and square of standard deviation is proportional to $Q$ which makes  $\frac{\sigma_{x}^2}{\mu_{x}^2} \propto \frac{1}{Q}$. Eventually, the standard deviation of the ratio distribution is inversely proportional to $Q$ while it does not affect the expected ratio value. Hence, as $Q$ increases PDFs of different transmitted ratios have lighter tails on both sides and thus have a smaller intersection area. This reduces the Gaussian noise, and receiver is able to demodulate the signals more successfully resulting in a drop in BER. Also ADMC algorithm achieves better error performance on larger $Q$ values, in other words where ISI becomes more significant compared to Gaussian noise, since ADMC was specifically designed to combat the effects of ISI. 
\begin{figure}[t]
    \centering
    \includegraphics[width=\linewidth]{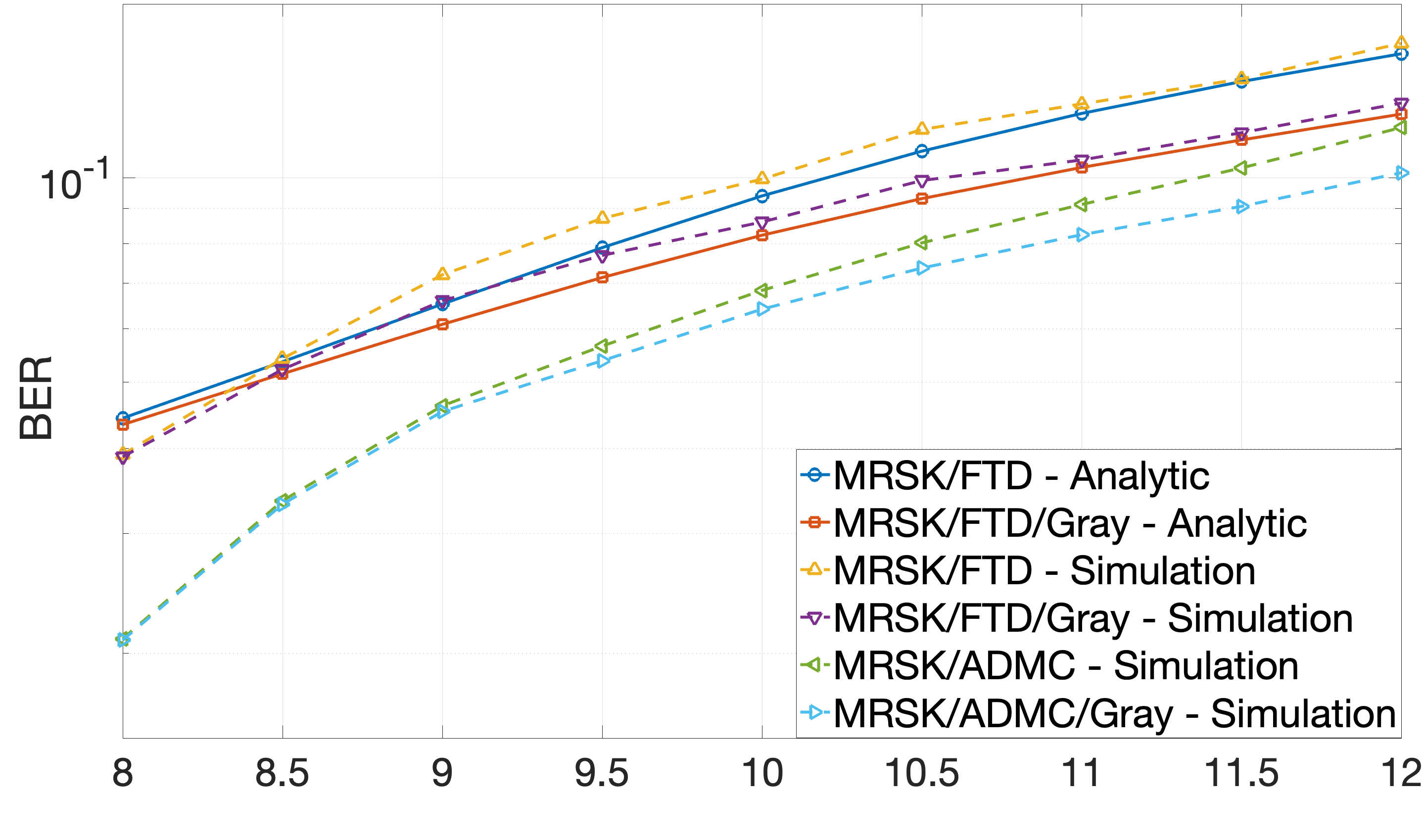}
    \caption{BER vs $d$ curves of MRSK.}
    \label{fig:MRSK_varying_d}
\end{figure}
\subsubsection{Distance Between Transmitter and Receiver}
BER curves are plotted in Fig. \ref{fig:MRSK_varying_d} to analyze the impact of the receiver-transmitter separation \( d \) by varying it within the range \( 8 - 12 \, \mu m \). Due to the diffusive nature of molecular communication, the molecular concentration diminishes with increasing \( d \), evident from \eqref{hit_probability}. As previously discussed, while CIR does not alter the expected value of the received molecular ratio, it inversely affects the standard deviation. Consequently, increasing \( d \) leads to a higher variance in the received molecular ratio, resulting in a wider spread in the PDFs of received ratios associated with different symbols. This spread corresponds to an increase in Gaussian noise, thereby causing a rise in BER as \( d \) grows. Moreover, we infer that same saturation effect explained in Section \ref{bit_time} causes binary and Gray-coded variants to diverge as $d$ increases since molecules are more dispersed further the receiver is located.

\begin{figure}
    \centering
    \includegraphics[width=\linewidth]{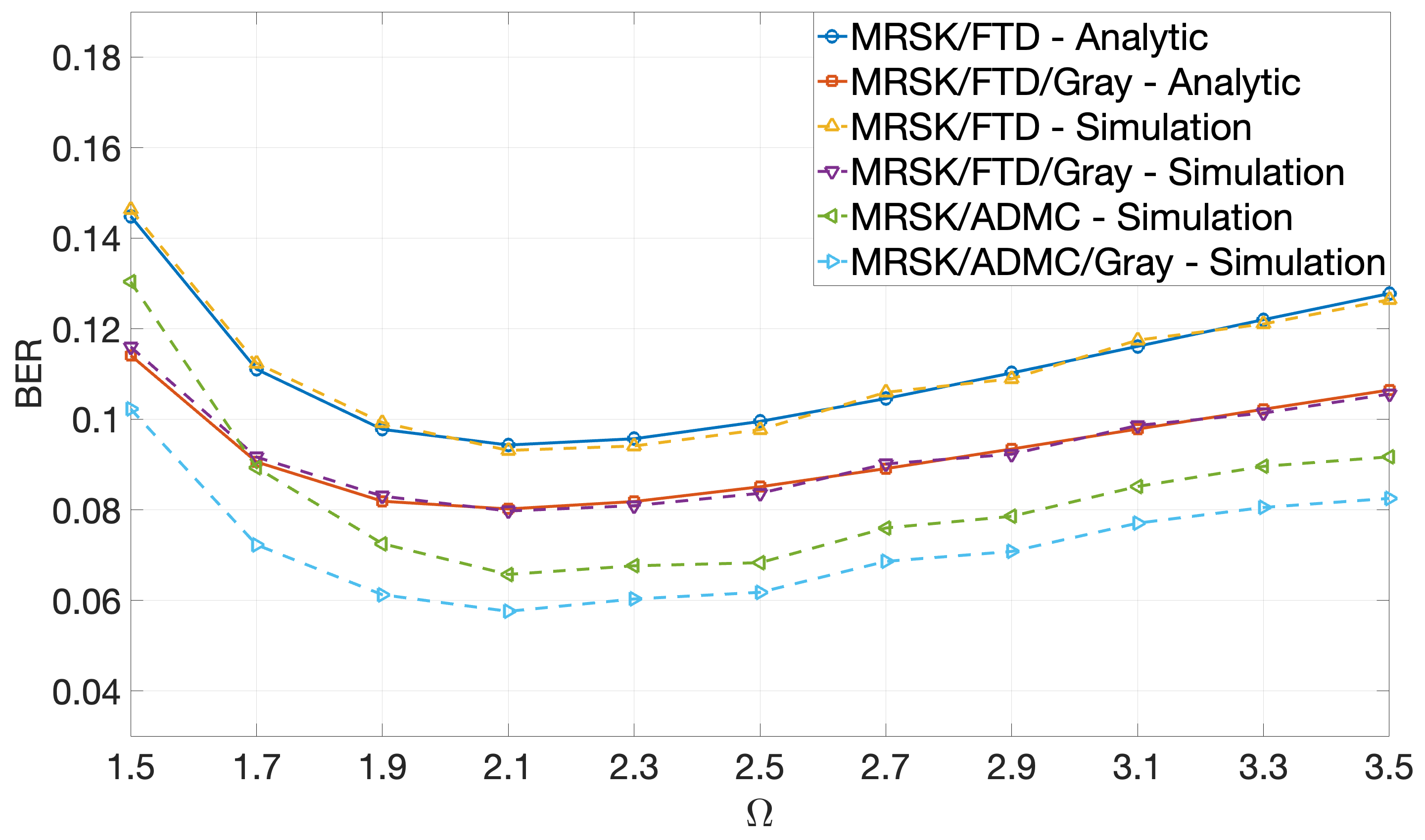}
    \caption{BER vs $\Omega$ curves of MRSK.}
    \label{fig:MRSK_varying_Omega}
\end{figure}

\subsubsection{Ratio Range}
\begin{figure}[t]
    \centering
    \includegraphics[width=\linewidth]{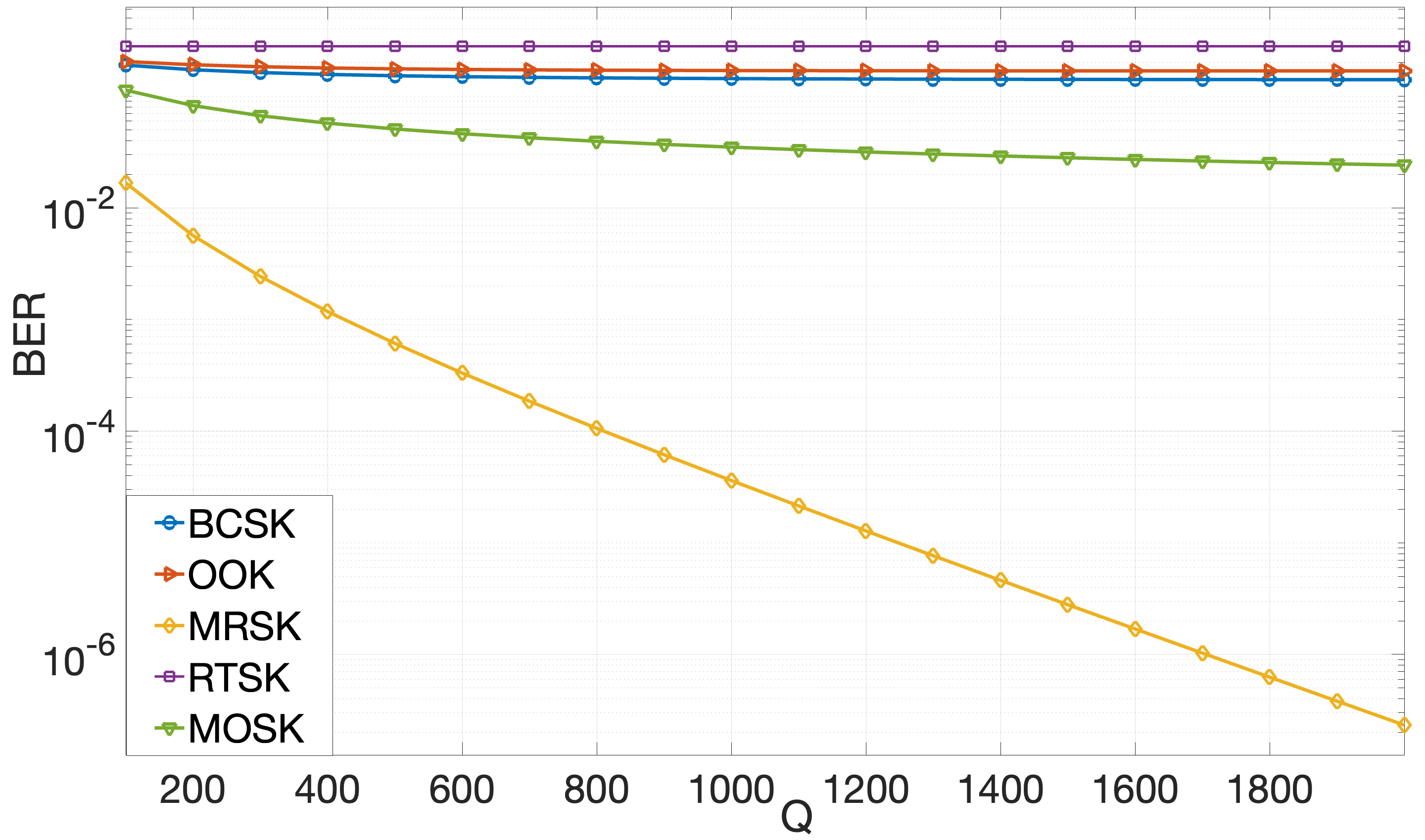}
    \caption{BER vs $Q$ curves of selected modulation schemes.}
    \label{fig:modulations_compared_varyingQ}
\end{figure}

Up until now, we have used $\Omega=e$ as the default value for the range of predefined ratios. As promised in Section \ref{MRSK}, we now analyze the effect of the parameter $\Omega$, which defines the range for the possible ratio values symbols can attain. Although it seems like a wider range of ratios might unconditionally decrease BER, it is not always the case for MRSK. We have plotted the BER performance of MRSK with both FTD and ADMC for varying $\Omega$ values in Fig. \ref{fig:modulations_compared_varyingQ}. From $\Omega=1.5$ to $\Omega=2.1$, the BER decreases across both FTD and ADMC because in this range very low number of molecules are transmitted and ISI is not of concern. So, the average number of molecules being transmitted reduces Gaussian noise, and thus decreases BER. After the minima at $\Omega=2.1$, however, high number of molecules saturates the receiver. As previously seen, MRSK suffers from ISI in this setting and error performance gradually deteriorates from this point onwards. 
\subsection{Comparison with Other Modulations}
Results of the previous subsections highlight that MRSK achieves optimal performance when $N=2$, $M=1$ for $t_b=0.5$. This configuration encodes bits using the ratio $e$ for bit 1 and $e^{-1}$ for bit 0. We now compare the BER performance of dual molecule MRSK with other commonly used SISO modulation schemes, including On-Off Keying (OOK) \cite{OOK}, Concentration Shift Keying (CSK) \cite{CSK}, Release Time Shift Keying (RTSK) \cite{RTSK} and Molecule Shift Keying (MOSK) \cite{MOSK}. FTD is utilized for MRSK due to both its independence of CIR and also to ensure fair grounds for comparison. Since analytical expression for BER of MRSK was verified by simulations, we evaluate the BER of each modulation analytically. Additionally, as BER decreases below $10^{-5}$ range, no error is observed even after the simulation is run numerous times, supporting the use of analytic calculations.

In order to ensure fairness, the same system model, featuring a point source transmitter and spherical absorbing receiver, used throughout the paper is applied here for modulation comparison with parameters $d=10\mu m$, $r=5\mu m$, $D=79.4\mu m^2/s$ , channel memory $L=5$. Gaussian distribution is assumed for the statistics of total number of received molecules, consistent with previous sections. $Q$ and $t_b$ are set to $Q=1000$ and $t_b=1 s$ unless they are swept across. 

For OOK, we adopt the methodology described in \cite{OOK_compared}. At the beginning of each symbol interval $(T_s=t_b)$, $Q$ number of molecules are released to transmit bit 1 while no molecules are released to transmit bit 0. On the receiver side, fixed threshold decoding is implemented with threshold $\mathcal{E}= \alpha Q$ where the value of $\alpha$ was optimized through a simple sweep search to $\alpha=0.78$. The error probability is calculated for each of the $2^5$ possible symbol sequences and then averaged to obtain BER. 

M-ary CSK in \cite{M-ary} is implemented for the BER analysis of CSK modulation with each symbol encoding 1 bit of information or BCSK as called by authors of \cite{M-ary}. This scheme uses two predefined molecule concentrations for transmission, with $Q_1$ set to $Q$ and each subsequent concentration scaled by a factor of $\Gamma$ ($Q_2= \Gamma Q_1$). $\Gamma$ is set to $\Gamma=2$ for number of molecules to be approximately equal to the MRSK case and $\rho=1$ is used for simplicity. Fixed thresholds were also employed for detection with thresholds $\mathcal{E}_i$ determined by the geometric mean of adjacent received molecule counts ($\mathcal{E}_i = \sqrt{z_i[k] \ z_{i-1}[k]}$), similar to FTD scheme implemented for MRSK. It is worth noting that our BER formulations differ slightly from those in \cite{M-ary}, as their study considered mobile MC, while our adaptation is for stationary MC scenarios.

In RTSK, information is encoded into the release timing of molecules, reminiscent of pulse position modulation. A diffusion-based molecular timing (DBMT) channel was assumed, with the same point-source transmitter and spherical absorbing receiver described earlier. Transmitted signals are detected based on the first arrival of the molecules released in time $t=\Delta$ after the beginning of each symbol interval. For a DBMT channel without flow, the received signal follows a Lévy distribution. We directly adopt the ML and linear detection methods outlined in \cite{RTSK_compared}. In  our implementation, we set $\Delta = \frac{t_b}{2}$ to ensure that  first-arriving molecules within a symbol interval do not interfere with the subsequent interval.
\begin{figure}[t]
    \centering
    \includegraphics[width=\linewidth]{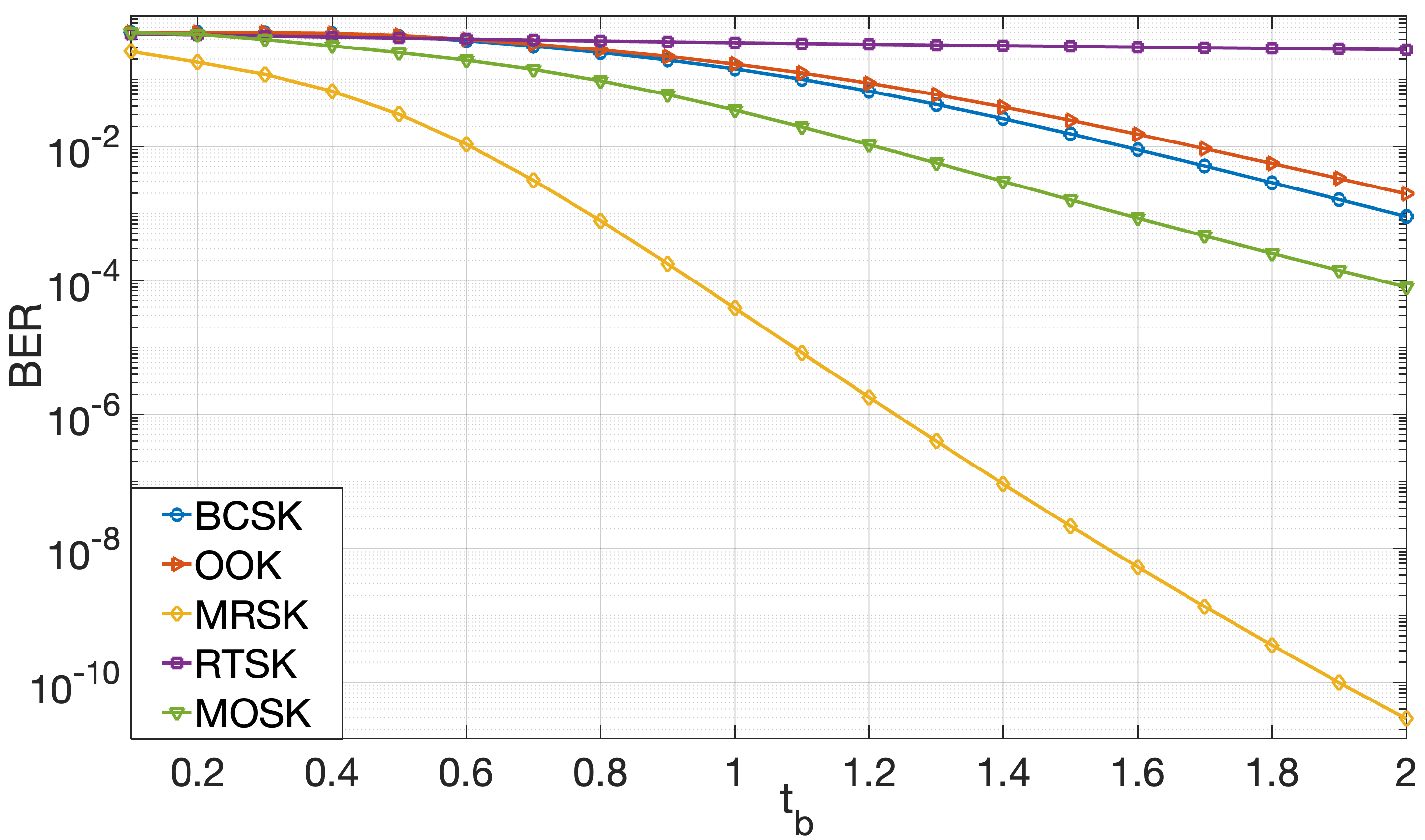}
    \caption{BER vs $t_b$ curves of selected modulations schemes.}
    \label{fig: modulations_compared_varying_tb}
\end{figure}

MoSK utilizes different molecule types to transmit information. This analysis considers SISO binary MoSK as presented in \cite{MOSK}. In this scheme, one of two molecules with similar diffusion coefficients encodes bit 0, while the other encodes bit 1. An error is assumed if neither or both of the received number of molecules are above a certain threshold as proposed in \cite{MOSK}. Again by a sweep search, optimal value of the threshold that minimizes the average BER is determined to be $\Lambda=0.34 Q$ for the given channel parameters. 

We first plot the BER performance of selected modulation schemes across a range of  $t_b$ values in Figure \ref{fig: modulations_compared_varying_tb}. CSK and OOK exhibit similar error patterns, which aligns with expectations since their underlying mechanisms are similar. MoSK performs better than single-molecule modulation schemes at the cost of using two distinct molecules. Notably, MRSK, equipped with FTD, significantly outperforms other modulation schemes. The BER gap widens as $t_b$ increases, indicating that ISI is the primary performance limiting factor for MRSK. However, it copes with the intrinsic Gaussian noise in MC systems more effectively than traditional schemes.

Finally, we examine the effect of $Q$ on selected modulation schemes. Increasing $Q$ essentially narrows the linewidth of Gaussian PDFs compared to their expected values and separates the PDFs further apart from each other which results in less intersection and, hence, lower probability of misdetection. Increasing the Q enhances the validity of approximating binomial arrivals as Gaussian \eqref{binomial_to_gaussian_arrival}. However, in DBMT channels, where detection is based on the first arrival of molecules, BER is independent of $Q$, as noted in \cite{RTSK_compared}. This aligns with our observation of a constant BER across the $Q$ range. Once again, MRSK demonstrates superior performance over other schemes within the given $Q$ range, with a significantly lower BER. Unlike bit time, the BER decreases exponentially with $Q$ as additional molecules eventually saturate the receiver and performance becomes ISI-limited.
\section{Conclusion}\label{Conclusion}
We introduce a novel modulation scheme, MRSK, which encodes the information into the ratios of molecule numbers, offering a substantial alternative to conventional MC modulation methods. A mathematical framework is developed to analyze the distribution of ratios of Gaussian random variables, enabling the precise derivation of the BER. Our findings reveal that binary modulation with minimal molecule types achieves optimal performance under moderate channel conditions while higher-order modulations or multiple molecule types effectively mitigate ISI in scenarios requiring shorter bit times. This adaptability makes MRSK particularly suitable for high-data-rate applications, as evidenced by its ability to leverage multiple molecule types to address such challenges. We also observe a convex relationship between ratio range and BER, underscoring the importance of parameter optimization to fully exploit the scheme's potentials. Comparisons with existing MC modulations consistently show that MRSK delivers superior BER performance, marking a significant advancement in MC.

Given MRSK's strong error performance and compatibility with biological systems, it offers a promising avenue for MC applications. Future research should focus on refining threshold optimization under varying channel conditions, investigating the scheme's resilience in diverse molecular networks, and exploring its integration into emerging biological and nanotechnology-based communication platforms.

\bibliographystyle{IEEEtran}
\bibliography{references.bib}

\begin{IEEEbiography}
    [{\includegraphics[width=1in, height=1.25in, clip, keepaspectratio]{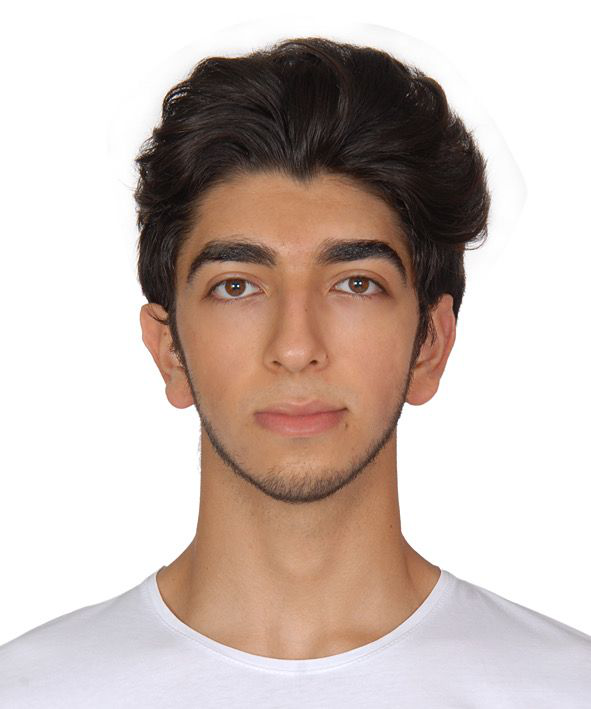}}]{Boran Aybak Kilic} completed his high school education at Kocaeli Enka Science and Technology High School, earning a bronze medal in the National Physics Olympiad. He is currently a third-year undergraduate student pursuing a double major in Electrical and Electronics Engineering and Physics at Bogazici University, Istanbul, Turkey. He serves as a research assistant at the Center for neXt-generation Communications (CXC) under the supervision of Prof. Ozgur B. Akan.
\end{IEEEbiography}
\vspace{1cm}
\begin{IEEEbiography}
[{\includegraphics[width=1in,height=1.25in,clip,keepaspectratio]{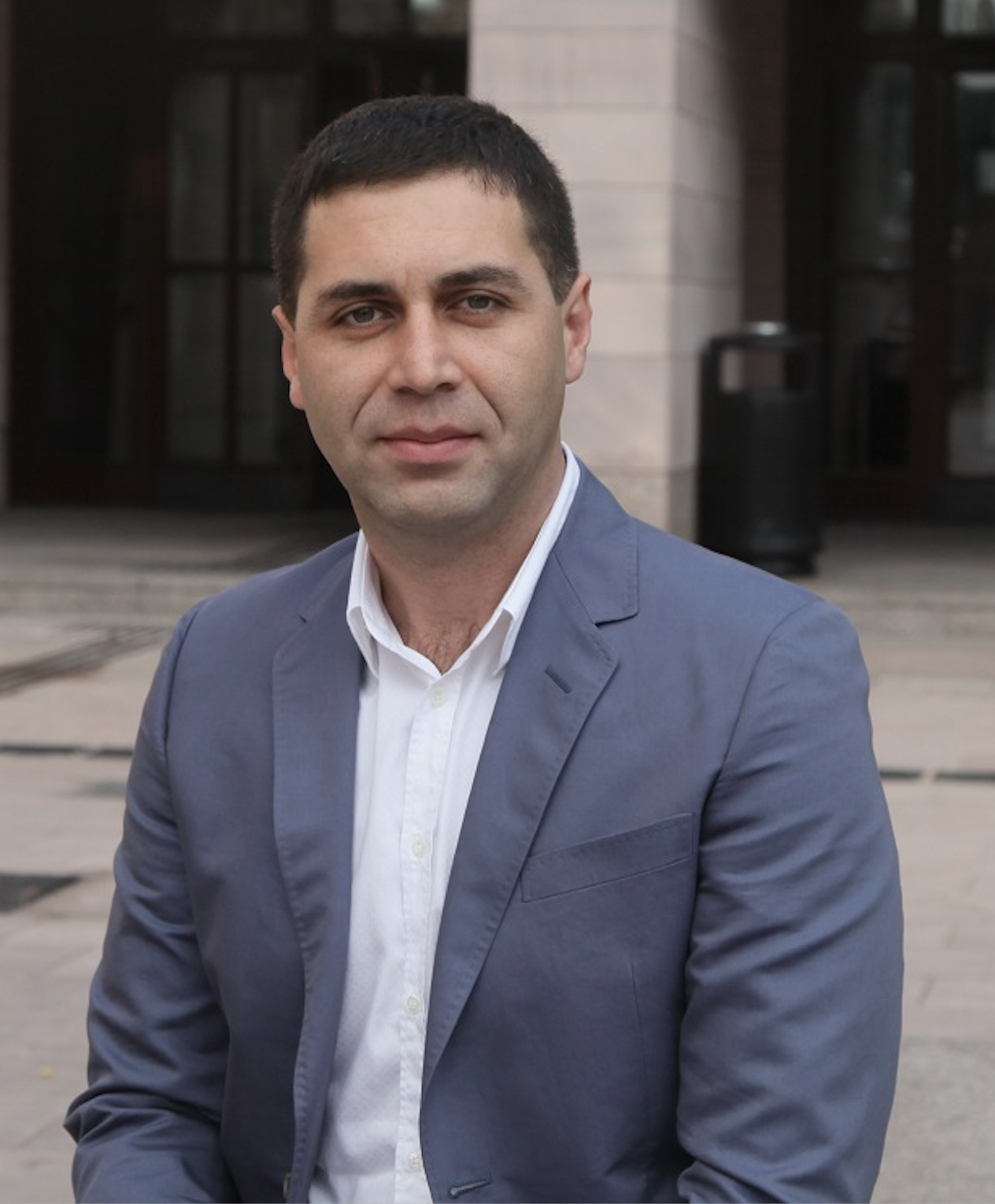}}]{Ozgur B. Akan (Fellow, IEEE)}
received the PhD from the School of Electrical and Computer Engineering Georgia Institute of Technology Atlanta, in 2004. He is currently the Head of Internet of Everything (IoE) Group, with the Department of Engineering, University of Cambridge, UK and the Director of Centre for neXt-generation Communications (CXC), Koç University, Turkey. His research interests include wireless, nano, and molecular communications and Internet of Everything.
\end{IEEEbiography}
\end{document}